# Multi-parametric sensitivity analysis of the band structure for tetrachiral acoustic metamaterials

Marco Lepidi[a], Andrea Bacigalupo[b,*]

[a]DICCA - Dipartimento di Ingegneria Civile, Chimica e Ambientale, Università di Genova, Via Montallegro 1, 16145 Genova (Italy)
[b]IMT School for Advanced Studies Lucca, Piazza S. Francesco 19, 55100 Lucca (Italy)

## Abstract

Tetrachiral materials are characterized by a cellular microstructure made by a periodic pattern of stiff rings and flexible ligaments. Their mechanical behaviour can be described by a planar lattice of rigid massive bodies and elastic massless beams. The periodic cell dynamics is governed by a monoatomic structural model, conveniently reduced to the only active degrees-of-freedom. The paper presents an explicit parametric description of the band structure governing the free propagation of elastic waves. By virtue of multiparametric perturbation techniques, sensitivity analyses are performed to achieve analytical asymptotic approximation of the dispersion functions. The parametric conditions for the existence of full band gaps in the low-frequency range are established. Furthermore, the band gap amplitude is analytically assessed in the admissible parameter range. In tetrachiral acoustic metamaterials, stop bands can be opened by the introduction of intra-ring resonators. Perturbation methods can efficiently deal with the consequent enlargement of the mechanical parameter space. Indeed high-accuracy parametric approximations are achieved for the band structure, enriched by the new optical branches related to the resonator frequencies. In particular, target stop bands in the metamaterial spectrum are analytically designed through the asymptotic solution of inverse spectral problems.

## 1. Introduction

Several theoretical and applied research fields are currently developing a renewed interest in the high mechanical performances of cellular and periodic materials. Consequently, their traditional role of efficient structural elements is undergoing a rapid evolution. This trend is also catalyzed by virtuous synergies with the recent extraordinary developments in parametric design, multi-scale modeling, computational techniques and multi-disciplinary meta-analyses. Within this scientific framework, advanced theoretical formulations and revolutionary manufacturing technologies contribute to offer solid prospects for the birth of new-generation materials, with superior mechanical properties and smart multi-field functionalities.

Within the specific context of solid and structural mechanics, a promising challenge focuses on exploiting the periodic microstructure and the marked anisotropy of some chiral or antichiral lattice materials to steer or stop elastic waves along particular directions. One of the simplest planar configuration realizing a chiral honeycomb consists of a regular microstructure made of stiff disks or rings, connected by a variable number of flexible ligaments [1–3]. The leading idea is that, within certain admissible ranges, the microstructural parameters can be employed as design variables to tailor the dispersion properties of the material. Among the possible technological advances, an appealing goal is the synthesis of highly-costumizable elastic media, suited to serve as mechanical guides or phononic filters for optical and acoustic signals.

Based on these motivations, considerable research attention has been devoted over the last decade to analyze and/or control the dispersion properties of different periodic microstructures with chiral characteristics. These analyses have been based on low-dimensional Lagrangian models [4–6], high-fidelity computational formulations [7–9], equivalent homogenized continua [10–14] and experimental prototypes [15]. A decisive boost to the research on this topic comes from the discovery that stop bands can be opened in the spectrum of a chiral solid through the introduction of local resonators. In periodic materials with chiral ring-ligament microstructures, local resonators can be realized by intra-ring masses elastically connected with the hosting rings. The microstructured material enriched with these auxiliary masses (inertial resonators) can be classified as an elastic or acoustic metamaterial (inertial metamaterial). Acoustic metamaterials have already attracted several studies focused on both direct and inverse spectral problems [16–19].

According to the simplest mathematical formulation, the *direct* spectral problem consists in determining the dispersion function $\omega(\mathbf{p}, \mathbf{k})$ in the Brillouin domain $\mathcal{B}$, spanned by the wavevector $\mathbf{k}$, for a certain periodic material, fully described by a given set $\mathbf{p}$ of microstructural parameters. The spectral design, instead, can be regarded as an *inverse* problem, consisting in determining which parameter set $\mathbf{p}$ realizes an unknown periodic material (if one exists), characterized by a desired dispersion function $\omega(\mathbf{p}, \mathbf{k})$. Clearly, the solution of any direct or inverse spectral problem can greatly benefit from the availabil-ity of an explicit form for the dispersion relation $\omega(\mathbf{p}, \mathbf{k})$, as an analytical function of the $(\mathbf{p}, \mathbf{k})$-variables. On the contrary, the eigenproblem governing the wave dispersion returns an implicit dispersion relation $F(\omega, \mathbf{p}, \mathbf{k}) = 0$, which is not explicitly invertible in the general case. Therefore, the numerical routine for the direct problem solution consists in the step-by-step construction of the frequency loci over a sufficiently fine discretization of the $\mathcal{B}$-domain, fixed the parameter set $\mathbf{p}$. Similarly, the typical attack to inverse problems is based on highly-demanding procedures of computational optimization [20–23].

Perturbation methods can represent an efficient alternative tool, suited to determine explicit – though asymptotically approximate – analytical expressions of the dispersion relation for periodic structures [24–27], as well as for their equivalent homogenized continua [28]. In [6], a multi-parameter perturbation strategy has been outlined to build up asymptotic approximations for the dispersion functions of Lagrangian lattice models. The strategy is based on including both the wavenumbers and the mechanical parameters in the small amplitude perturbation vector $\boldsymbol{\mu}' = (\mathbf{p}', \mathbf{k}')$, spanning a small-radius multi-dimensional hyper-sphere centered at a suited reference point $\boldsymbol{\mu}^\circ$ of the multi-parameter space. By virtue of a recursive formulation, the perturbation equations governing the direct spectral problem up to the desired approximation order have been derived for a generic lattice, independently of the dimensions of the Lagrangian model and the parameter space. The equation solutions, consisting of the multi-parametric sensitivities of the dispersion functions, have been verified to well-approximate the spectrum of the anti-tetrachiral material, along specific directions of the $\mathcal{B}$-domain. Furthermore, the same perturbation strategy has been successfully applied to attack a basic inverse spectral problem [29].

The purpose of the present work is to upgrade the perturbation strategy, in order to govern advanced direct and inverse problems. From the mathematical viewpoint, the perturbation strategy outlined in [6] is here improved by the closed form solution of the perturbation equations. This novel achievement allows the direct formal assessment of all the frequency sensitivities as analytical functions of the multi-parameter perturbation (for the general case $\partial F/\partial \omega \neq 0$ in $\omega(\boldsymbol{\mu}^\circ)$). From the methodological viewpoint, the effectiveness and versatility of the perturbation strategy are here leveraged to address two unexplored tasks related to direct and inverse spectral problems. The tetrachiral microstructure is chosen as benchmark material typology. The first task concerns the systematic employment of the multi-parameter perturbation technique for the asymptotic approximation of the dispersion surfaces over the *entire* irreducible Brillouin zone. In this respect, a complementary issue regards the search for full band gaps in the material spectrum. Indeed, the availability of parametric dispersion functions strongly facilitates the establishment of mechanical conditions for the existence of stop bands, as well as for the assessment of their amplitude. The second task regards the novel application of the multi-parameter perturbation technique to parametrically approximate the dispersion curves of the acoustic *metamaterial*, whose cellular microstructure is characterized by a higher model dimension and a larger parameter space. In partic-
ular, the challenging point consists in preserving the accuracy of the asymptotic approximation in a denser dispersion spectrum, enriched by the resonator frequencies. The final substantial purpose is the parametric design of the resonator properties, in order to govern the opening of a stop band in the material spectrum, by simply assigning the desired bandwidth, centered around a frequency of interest.

The paper is organized as follows. First, a Lagrangian structural model is formulated to describe the free undamped dynamics of the periodic cell for the tetrachiral material and metamaterial (Section 2). Second, the Floquet-Bloch theory is followed to state the wave propagation problem (Section 3), and the multi-parameter perturbation strategy of solution is presented, with some attention paid to the novel algorithmic developments (Paragraph 3.1). Therefore, the exact and asymptotically approximate dispersion functions are presented for the tetrachiral material (Section 4) and metamaterial (Section 5). Different technical issues are analyzed, including the location of the unperturbed points in the parameter space, the dimension of the perturbation vector, the accuracy and validity of the asymptotic approximation, the suitability of a certain parameter ordering. In parallel, several aspects of mechanical interest are discussed, concerning the occurrence of multiple frequencies, the conditions for the existence of low-frequency stop bands, the parametric assessment of the stop bandwidth, the evaluation and improvement of the resonator effects (Paragraphs 4.1 and 5.1). In particular, design procedures for the inertial resonator parameters are sketched, in order to open a desired stop band in the material spectrum (Paragraph 5.2). Conclusive remarks are finally pointed out.

## 2. Beam lattice model

The two-dimensional geometric pattern of the tetrachiral material is based on a regular, periodic tessellation of the infinite Euclidean plane with square cells (Figure 1a). The mechanical behaviour of each elementary cell is characterized by an internal centro-symmetric structure, or *microstructure*, including a central circular ring connected to four tangent ligaments organized according to a chiral scheme (Figure 1b). By virtue of the chirality, the linear ring-ligament coupling may activate the so-called *rolling up* deformation mechanism, accountable for the distinctive auxeticity connoting the elastic macroscopic response of the material [3, 11, 30–34].

A synthetic but accurate approach to formulate a low-dimensional Lagrangian model of the tetrachiral material can be based on a beam lattice formulation. According to this idea, the circular ring is supposed heavy and sufficiently stiff to be modeled as a massive rigid body. The ring mass $M$ and rotational inertia $J$ can be freely assigned, by independently setting the mean diameter $D$, the annular width and the material density. The varied configuration of the rigid body is described by the three planar displacements (in-plane translations $U_1, V_1$ and rotation $\theta_1$) of the configurational node ① located at the ring centroid (Figure 1c). The four identical ligaments are assumed sufficiently light and flexible to be described by linear, extensible, unshearable and massless beams, with elastic and geomet-ric properties defining the extensional $EA$ and flexural rigidity $EI$. The natural length of the beams is $L=H\cos\beta$, where $H$ is the cell side length and $\beta=\arcsin(D/H)$ is the ligament inclination angle with respect to the mesh lines connecting the ring centres. The ring-beam connections are considered perfectly-rigid joints. Due to the geometric periodicity, the cell boundary crosses the midspan of each ligament. Consequently, the varied configuration of the $i$-th beam is fully described by the rigid motion of the ligament-ring connection, at the one end, and the three planar displacements $(U_i, V_i, \theta_i)$ of the $i$-th configurational node ($i =$②,...,⑤) located at the cell boundary, at the other end.

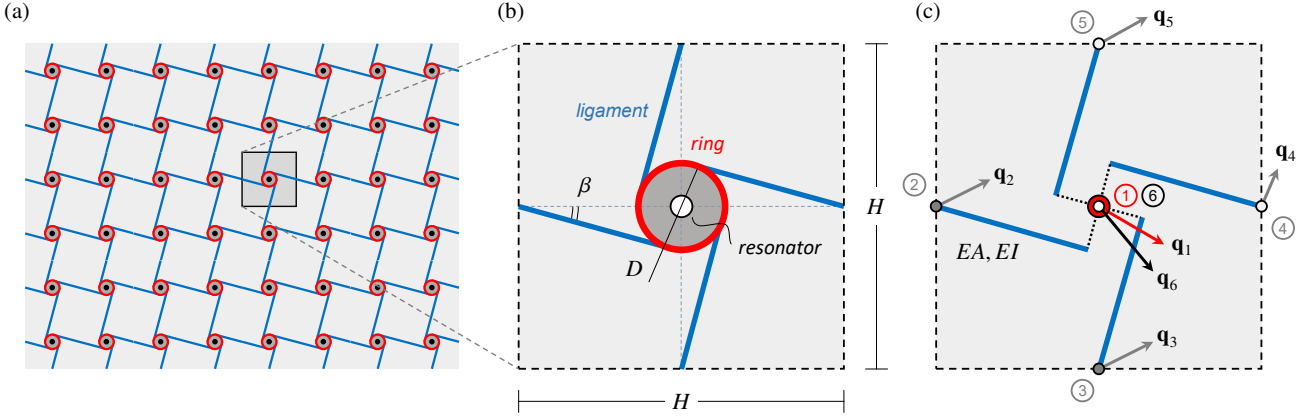

Figure 1: Tetrachiral metamaterial (a) repetitive planar pattern, (b) centro-symmetric microstructure of the periodic cell, (c) beam lattice model (black dotted lines represent rigid-end offsets connecting the beam-ring connections with the ring centroid).

The acoustic metamaterial can be obtained by adding local resonators consisting in intra-ring circular masses, which are co-centered and elastically coupled with the hosting rings. Each resonator is modeled as an undamped linear oscillator with mass $M_r$ and rotational inertia $J_\theta$. The stiffness of the ring-resonator coupling is ideally provided by a pair of translational springs and a rotational spring with elastic constants $K_r$ and $K_\theta$, respectively. Therefore, the three circular frequencies of the free-standing oscillator are $\Omega_r^2 = K_r/M_r$ (double frequency) and $\Omega_\theta^2 = K_\theta/J_\theta$. The oscillator dynamics is described by the three planar displacements $(U_6, V_6, \theta_6)$ of the configurational node ⑥. It is worth remarking that, in the reference configuration, the configurational nodes ① and ⑥ share the same position at the ring centroid. The ring-resonator location sharing in beam lattice models fully complies with the general concept of acoustic metamaterials with local resonances [35, 36], that is, periodic lattices characterized by material points singularly (and locally) coupled with tuned mass-spring oscillators.

Denoting $\Omega_c$ a known circular frequency serving as auxiliary dimensional reference, the inertial, elastic and geometric properties of the cell microstructure are described by the minimal set **p** of independent nondimensional parameters

$$\delta=\frac{D}{H}, \quad \varrho^2=\frac{I}{AL^2}, \quad \chi^2=\frac{J}{MH^2}, \quad \omega_c^2=\frac{EA}{MH\Omega_c^2} \quad (1)$$

where $\delta$ expresses the spatial density of the rings, measuring also the material mass density at the macroscopic scale.

The inverse of the nondimensional radius of gyration $\varrho$ accounts for the beam slenderness. The $\chi^2$-parameter describes the rotational-to-translational mass ratio of the rings, while $\omega_c$ is a nondimensional normalization frequency, which can be assumed to be unitary in the following without loss of generality.

Additional mechanical parameters are necessary to describe the inertial resonators of the metamaterial. The resonator masses and frequencies are taken as principal parameters, and the stiffnesses as dependent quantities (according to the rule $K_r = \Omega_r^2 M_r$ and $K_\theta = \Omega_\theta^2 J_\theta$). Therefore, the minimal set of independent parameters **p** describing the metamaterial cell must be enriched by the nondimensional quantities

$$\gamma=\frac{\Omega_r}{\omega_c\Omega_c}, \quad \alpha^2=\frac{M_r}{M}, \quad \gamma_\theta=\frac{\Omega_\theta}{\omega_c\Omega_c}, \quad \chi_r^2=\frac{J_\theta}{M_rH^2} \quad (2)$$

where, among the others, the nondimensional resonator frequency $\gamma$ and resonator-to-ring mass ratio $\alpha^2$ are the most relevant parameters.

### 2.1. Equation of motion

According to the mechanical assumptions, the linear dynamics of the periodic cell is governed by a multi-degrees-of-freedom model, referred to six configuration nodes. The actual configuration of the $i$-th node is described by the nondimensional displacement vector $\mathbf{q}_i = (u_i, v_i, \theta_i)$, where the variables

$$u_i = \frac{U_i}{H}, \qquad v_i = \frac{V_i}{H} \quad (3)$$

Consequently, the cell configuration is described by a 15-by-1 displacement vector $\mathbf{q} = (\mathbf{q}_1, ..., \mathbf{q}_5)$ for the tetrachiral material, or a 18-by-1 displacement vector $\mathbf{q} = (\mathbf{q}_1, ..., \mathbf{q}_6)$ for the tetrachiral metamaterial.

Employing the direct stiffness method to govern the equilibrium, the undamped free response of the Lagrangian model is governed by the ordinary differential equations of motion

$$\mathbf{M}(\mathbf{p})\ddot{\mathbf{q}} + \mathbf{K}(\mathbf{p})\mathbf{q} = \mathbf{f} \quad (4)$$

where the dot denotes differentiation with respect to the nondimensional time $\tau = \Omega_c t$, while $\mathbf{M}(\mathbf{p})$ and $\mathbf{K}(\mathbf{p})$ are the parameter-dependent mass and stiffness matrices. Finally, the vector $\mathbf{f}$ collects the elastic forces exerted by the adjacent cells to the boundary nodes.

The configuration vector can be partitioned in the form $\mathbf{q} = (\mathbf{q}_a, \mathbf{q}_p)$ by properly distinguishing

- the *active* displacements $\mathbf{q}_a$ of the massive central nodes, where inertial forces may develop
- the *passive* displacements $\mathbf{q}_p$ of the massless boundary nodes, where only static forces may act.

To specify, the *active* displacement vector is $\mathbf{q}_a = (\mathbf{q}_1, \mathbf{q}_6)$ for the metamaterial (but $\mathbf{q}_a = \mathbf{q}_1$ in the absence of inertial resonators), while the *passive* displacement vector is $\mathbf{q}_p = (\mathbf{q}_2, ..., \mathbf{q}_5)$. According to the displacement partition and dropping the matrix dependence on $\mathbf{p}$, the equations of motion read

$$\begin{bmatrix} \mathbf{M}_a & \mathbf{O} \\ \mathbf{O} & \mathbf{O} \end{bmatrix} \begin{pmatrix} \ddot{\mathbf{q}}_a \\ \ddot{\mathbf{q}}_p \end{pmatrix} + \begin{bmatrix} \mathbf{K}_{aa} & \mathbf{K}_{ap} \\ \mathbf{K}_{pa} & \mathbf{K}_{pp} \end{bmatrix} \begin{pmatrix} \mathbf{q}_a \\ \mathbf{q}_p \end{pmatrix} = \begin{pmatrix} \mathbf{0} \\ \mathbf{f}_p \end{pmatrix} \quad (5)$$

where $\mathbf{O}$ stands for empty matrices. The lower equation expresses the quasi-static equilibrium regulating the passive displacements of the boundary nodes.

Therefore, the elastic wave propagation through the cellular material can be analyzed by exploiting the structural periodicity. In particular, the Floquet-Bloch boundary conditions can be imposed on the displacements and forces of the boundary nodes, in order to account for the planar waves propagating across adjacent cells. Denoting $k_1$ and $k_2$ the wavenumbers of the horizontally and vertically propagating waves, respectively, the nondimensional wavevector $\mathbf{b} = (\beta_1, \beta_2)$ can be defined, where $\beta_1 = k_1 H$ and $\beta_2 = k_2 H$. The $\mathbf{b}$-vector spans the first Brillouin zone $\mathcal{B} = [-\pi, \pi] \times [-\pi, \pi]$.

Moreover, a suited reduction of the model dimension can be achieved by applying a classic quasi-static condensation to the *passive* displacements. This procedure univocally relates the passive to the active displacements according to a linear law, without any further approximation. It may be worth remarking that no algebraic adjustments are required to apply the condensation procedure to the metamaterial, since the resonators are not coupled with the boundary nodes. Finally, the free wave dynamics is governed by the reduced equation

$$\mathbf{M}_a(\mathbf{p}) \ddot{\mathbf{q}}_a + \mathbf{K}_a(\mathbf{p}, \mathbf{b}) \mathbf{q}_a = \mathbf{0} \quad (6)$$

where $\mathbf{K}_a(\mathbf{p}, \mathbf{b})$ is a $\mathbf{b}$-dependent generalization of the condensed stiffness matrix, with Hermitian properties. The parametric matrices $\mathbf{M}_a(\mathbf{p})$, $\mathbf{K}_a(\mathbf{p}, \mathbf{b})$ are reported in the AppendixA for the tetrachiral material and metamaterial.

## 3. Band structure

The wave equation (6) can be tackled by imposing the harmonic mono-frequent solution $\mathbf{q}_a = \boldsymbol{\psi}_a \exp(i\omega\tau)$, where $\omega = \Omega/\Omega_c$ and $\Omega$ are the unknown nondimensional and dimensional wave frequency, respectively. Therefore, eliminating the dependence on time, a linear eigenproblem can be obtained in the standard form

$$(\mathbf{H}(\mathbf{p}, \mathbf{b}) - \lambda \mathbf{I})\boldsymbol{\phi}_a = \mathbf{0} \quad (7)$$

where $\lambda = \omega^2$ and $\mathbf{H}(\mathbf{p}, \mathbf{b}) = \mathbf{Q}(\mathbf{p})^{-\top} \mathbf{K}_a(\mathbf{p}, \mathbf{b}) \mathbf{Q}(\mathbf{p})^{-1}$, with $\mathbf{Q}(\mathbf{p})$ following from the unique decomposition of the diagonal matrix $\mathbf{M}(\mathbf{p}) = \mathbf{Q}(\mathbf{p})^\top \mathbf{Q}(\mathbf{p})$. Denoting $F(\lambda, \mathbf{p}, \mathbf{b})$ the characteristic polynomial of the matrix $\mathbf{H}(\mathbf{p}, \mathbf{b})$, the eigenvalues $\lambda(\mathbf{p}, \mathbf{b})$ are the real-valued roots of the characteristic equation $F(\lambda, \mathbf{p}, \mathbf{b}) = 0$. The corresponding complex-valued eigenvectors $\boldsymbol{\psi}_a(\mathbf{p}, \mathbf{b}) = \mathbf{Q}(\mathbf{p})^{-1}\boldsymbol{\phi}_a(\mathbf{p}, \mathbf{b})$ represent the polarization mode of the $\omega(\mathbf{p}, \mathbf{b})$-monofrequent propagating wave. Depending on the different model dimension, the tetrachiral material and metamaterial possess three and six eigenpairs $(\lambda, \boldsymbol{\phi}_a)$, respectively. The dispersion functions $\omega(\mathbf{b})$ over the subdomain $\mathcal{B}_1 \subset \mathcal{B}$, bounded by the minimal contour of the first irreducible Brillouin zone, fully characterize the Floquet-Bloch spectrum (or band structure) of the material described by certain mechanical parameters $\mathbf{p}$ [37].

The parametric analysis of the material band structure is usually performed numerically, by carrying out the dispersion surfaces (over the first irreducible Brillouin zone $\mathcal{B}_1$) or the dispersion curves (along the closed boundary $\partial \mathcal{B}_1$ of the $\mathcal{B}_1$-zone). Indeed, in the absence of closed-form eigensolutions, the typical computational approach consists in a brute-force treatment of the implicit dispersion relation $F(\lambda, \mathbf{p}, \mathbf{b}) = 0$. Basically, the simplest procedure is based on the step-by-step construction of the eigenvalue (or frequency) loci over a sufficiently fine discretization of the $\mathbf{b}$-space, fixed a certain $\mathbf{p}$-value. Therefore, the loci construction is cycled until the whole parameter range of interest is spanned by gradual $\mathbf{p}$-updates.

### 3.1. Multi-parameter perturbation method

A proper alternative to the numerical description of the band structure may consist in seeking for an explicit – though approximate – analytic form of the dispersion relation $\lambda = f(\mathbf{p}, \mathbf{b})$. To this purpose, perturbation methods turn out to be efficient mathematical tools, suited to provide asymptotic approximations of the dispersion functions with fine quantitative accuracy and wide-range validity in the parameter space.

Following a multi-parameter perturbation method (MPPM), suited for high-dimensional, multi-parametric Lagrangian models of periodic materials with dense spectra [6, 38, 39], the extended parameter vector $\boldsymbol{\mu} = (\mathbf{p}, \mathbf{b})$ can be introduced. Therefore, *all* the mechanical parameters and the wavenumbers are simultaneously scaled according to their smallness by assigning the tentative ordering

$$\boldsymbol{\mu}(\epsilon) = \boldsymbol{\mu}^\circ + \epsilon \boldsymbol{\mu}' \quad (8)$$

where the small nondimensional parameter $\epsilon \ll 1$ has a mere algorithmic role and must not be interpreted as a physical variable, according to the well-established literature of multi-parameter perturbation techniques (e.g. the Multiple Scale Method). If necessary, higher order terms (namely $\epsilon^n \boldsymbol{\mu}^{(n)}$ with $n > 1$) can be added.

As long as the matrix $\mathbf{H}(\boldsymbol{\mu}^\circ)$ is Hermitian, it does not admit defective eigenvalues. Therefore, the generic eigenvalue satisfying the characteristic equation can be postulated to admit the asymptotic expansion in integer $\epsilon$-power series

$$\lambda(\epsilon) = \lambda^\circ + \epsilon \lambda' + \epsilon^2 \lambda'' + ... + \epsilon^n \lambda^{(n)} + ... \quad (9)$$

where the $(\boldsymbol{\mu}^\circ, \boldsymbol{\mu}')$-dependent coefficient $\lambda^{(n)}$ of the series can be regarded as the $n$-th $\epsilon$-derivative (evaluated at $\epsilon = 0$ and divided by the factorial $n!$) of the exact but implicit eigenvalue function $F(\lambda, \boldsymbol{\mu}(\epsilon)) = 0$. In the parameter space, $\lambda^{(n)}$ represents the directional derivative in the $\boldsymbol{\mu}'$-direction (evaluated in $\boldsymbol{\mu} = \boldsymbol{\mu}^\circ$), and is known as the $n$-th eigenvalue sensitivity, or *eigensensitivity*.

According to the classic perturbation theory, the series expansion $\lambda(\epsilon)$ is expected to provide a good *local* approximation, that is, to well-match the exact eigenvalue in the neighborhood of the reference parameter set $\boldsymbol{\mu}^\circ = (\mathbf{p}^\circ, \mathbf{b}^\circ)$, for a generic (small) multi-parametric perturbation $\boldsymbol{\mu}' = (\mathbf{p}', \mathbf{b}')$. In the absence of singularities, the approximation accuracy can be improved and extended by adding terms to the $\lambda(\epsilon)$-series. Thus, performing a *local multi-parameter sensitivity analysis* of the band structure consists in analytically determining all the local eigensensitivities $\lambda^{(n)}(\boldsymbol{\mu}^\circ, \boldsymbol{\mu}')$ up to the desired $n$-th order.

Two brief remarks are worth being pointed out to clarify the features of this perturbation scheme. First, small perturbation of the mechanical parameters or wavenumbers can be treated indifferently, since both $\mathbf{p}'$ and $\mathbf{b}'$ represent particular directions in the extended parameter space. Second, the reference point $\boldsymbol{\mu}^\circ = (\mathbf{p}^\circ, \mathbf{b}^\circ)$ in the parameter space plays a mere algorithmic role as starting point for the local sensitivity analysis. Consequently, the unperturbed microstructure described to the parameters $\mathbf{p}^\circ$ is actually a mathematical abstraction, which may even not correspond to a physical realization of the material.

Introducing the parameter ordering $\boldsymbol{\mu}(\epsilon)$ and the eigenvalue expansion $\lambda(\epsilon)$, the characteristic function $G(\epsilon) = F(\lambda(\epsilon), \boldsymbol{\mu}(\epsilon))$ can be expanded in $\epsilon$-series

$$G(\epsilon) = G^\circ + \epsilon G' + \epsilon^2 \frac{G''}{2!} + ... + \epsilon^n \frac{G^{(n)}}{n!} + ... \quad (10)$$

where the $n$-th coefficient requires to determine the $n$-th $\epsilon$-derivative of a multi-variable composite function (in $\epsilon = 0$). After some manipulations to work with the multi-variable case, the Scott-Faà di Bruno's formula ([40]) can be applied to deal with the inner series functions $\lambda(\epsilon), \boldsymbol{\mu}(\epsilon)$, yielding

$$G^{(n)} = \sum_{S_0(h,k)} \sum_{|p|=k} \frac{n! \mathbf{F}^{(h,|p|)}}{(h+k)!} \left[\mathbf{r}_{hp}^{[n]}\right]_{\boldsymbol{\mu}^\circ \Rightarrow \mathbf{0} \| \lambda^\circ \Rightarrow 0} \quad (11)$$

where the $(h,k)$-index set $S(h,k) = (h, k \in [0, h+k = n])$ and $S_0(h,k) = S(h,k) - (0,0)$. The instruction of mutual exclusion $\boldsymbol{\mu}^\circ \Rightarrow \mathbf{0} \| \lambda^\circ \Rightarrow 0$ requires zeroing either $\boldsymbol{\mu}^\circ$ or $\lambda^\circ$ after formal $\epsilon$-differentiation.

Defining $\ell$ the dimension of the extended parameter vector $\boldsymbol{\mu} = (\mu_1, ..., \mu_\ell)$, the partial derivatives for generic $h, k$ read

$$\mathbf{F}^{(h,|p|)} = \frac{\partial^h}{\partial \lambda^h} \frac{\partial^{|p|} F(\lambda, \mu_{p_1}, ..., \mu_{p_k})}{\partial \mu_{p_1} ... \partial \mu_{p_k}}, \quad (12)$$

while the multi-parametric recursive term $\mathbf{r}_{hp}^{[n]}$ is defined by the non-differential, polynomial function

$$\mathbf{r}_{hp}^{[n]} = \sum_{j=1}^{n} \left(\frac{a_{jhpn}}{n}\right)\left(\frac{\lambda^{(j)}}{\lambda^\circ} + \sum_{i=1}^{\ell} \frac{\mu_i^{(j)}}{\mu_i^\circ}\right) \mathbf{r}_{hp}^{[n-j]} \quad (13)$$

with initialization $\mathbf{r}_{hp}^{[0]} = (\lambda^\circ)^h (\mu^\circ)^p$ and $a_{jhpn} = j(h+|p|+1) - n$. The improvement with respect to classical formulas is that the equation (11) holds for a generic $\ell$-number of perturbation parameters, without limitations in the $n$-order of the approximation. The generic high-order parameter perturbation $\boldsymbol{\mu}(\epsilon) = \boldsymbol{\mu}^\circ + \epsilon \boldsymbol{\mu}' + ... + \epsilon^n \boldsymbol{\mu}^{(n)}$ is also taken into account. Furthermore, a major mathematical advantage is that this formulation leverages analytical $\epsilon$-derivatives to by-pass the applicative efforts of differentiating the composite function $F(\lambda(\epsilon), \boldsymbol{\mu}(\epsilon))$ with respect to the inner variable.

The perturbation method requires the characteristic equation to be asymptotically satisfied. Thus, an ordered chain of $n$ perturbation equations is stated by zeroing each $\epsilon^{(n)}$-order coefficient $G^{(n)}$. The solution algorithm depends on the individual algebraic multiplicity $m^\circ$ of the eigenvalues $\lambda^\circ$ satisfying the zeroth-order equation $G^\circ = F(\lambda^\circ, \boldsymbol{\mu}^\circ) = 0$. The theoretical background is fully illustrated in [6], where two fundamental algorithmic cases ($m^\circ = 1$ and $m^\circ = 2$) are sketched, with focus on the mathematical treatment of the low-order indeterminacies which rise up for $m^\circ > 1$.

As original development with respect to this background, the analytical formula for the unique solution of the $n$-th order perturbation equation has been determined (exactly). Indeed, in the simplest case of a single generating eigenvalue ($m^\circ = 1$, which also implies $F^{(1,0)} \neq 0$), the explicit multi-parametric formula for the $n$-th order eigensensitivity $\lambda^{(n)}$ reads

$$\lambda^{(n)} = -\frac{1}{F^{(1,0)}} \sum_{S_0(h,k)} \sum_{|p|=k} \frac{\mathbf{F}^{(h,|p|)}}{(h+k)!} \left[\mathbf{s}_{hp}^{[n]}\right]_{\boldsymbol{\mu}^\circ \Rightarrow \mathbf{0} \| \lambda^\circ \Rightarrow 0} \quad (14)$$

where $n > 1$ and the multi-parametric recursive term $\mathbf{s}_{hp}^{[n]}$ is defined by the polynomial function

$$\mathbf{s}_{hp}^{[n]} = (h+|p|)\left(\sum_{i=1}^{\ell} \frac{\mu_i^{(n)}}{\mu_i^\circ}\right)(\lambda^\circ)^h (\mu^\circ)^p + \\ + \sum_{j=1}^{n-1}\left(\frac{a_{jhpn}}{n}\right)\left(\frac{\lambda^{(j)}}{\lambda^\circ} + \sum_{i=1}^{\ell} \frac{\mu_i^{(j)}}{\mu_i^\circ}\right) \mathbf{s}_{hp}^{[n-j]} \quad (15)$$

which has to be initialized with $\mathbf{s}_{hp}^{[0]} = (\lambda^\circ)^h (\mu^\circ)^p$. After all the calculations, the $\epsilon$-parameter concludes its auxiliary role and can be completely re-absorbed to obtain the eigenvalue approximation (9) in the form $\lambda = f(\boldsymbol{\mu})$. In the practice, the parameter ordering (8) must be inverted (so that the relation $\boldsymbol{\mu}' = \epsilon^{-1}(\boldsymbol{\mu} - \boldsymbol{\mu}^\circ)$ is obtained) and substituted in each sensitivity.

In summary, the equation (14) allows to analytically build up the asymptotic multi-parameter approximation of the dispersion functions $\lambda = f(\mathbf{p}, \mathbf{b})$ up to the desired $n$-th order, by simply evaluating the partial $(\lambda, \mathbf{p}, \mathbf{b})$-derivatives of the characteristic function $F(\lambda, \mathbf{p}, \mathbf{b})$. Apart pathological situations, the asymptotic approximation is expected to be locally accurate, that is, to well-match the exact solution within a small region (typically, the sphere with $\epsilon$-radius) centered in the $(\mathbf{p}^\circ, \mathbf{b}^\circ)$-point of the extended parameter space.

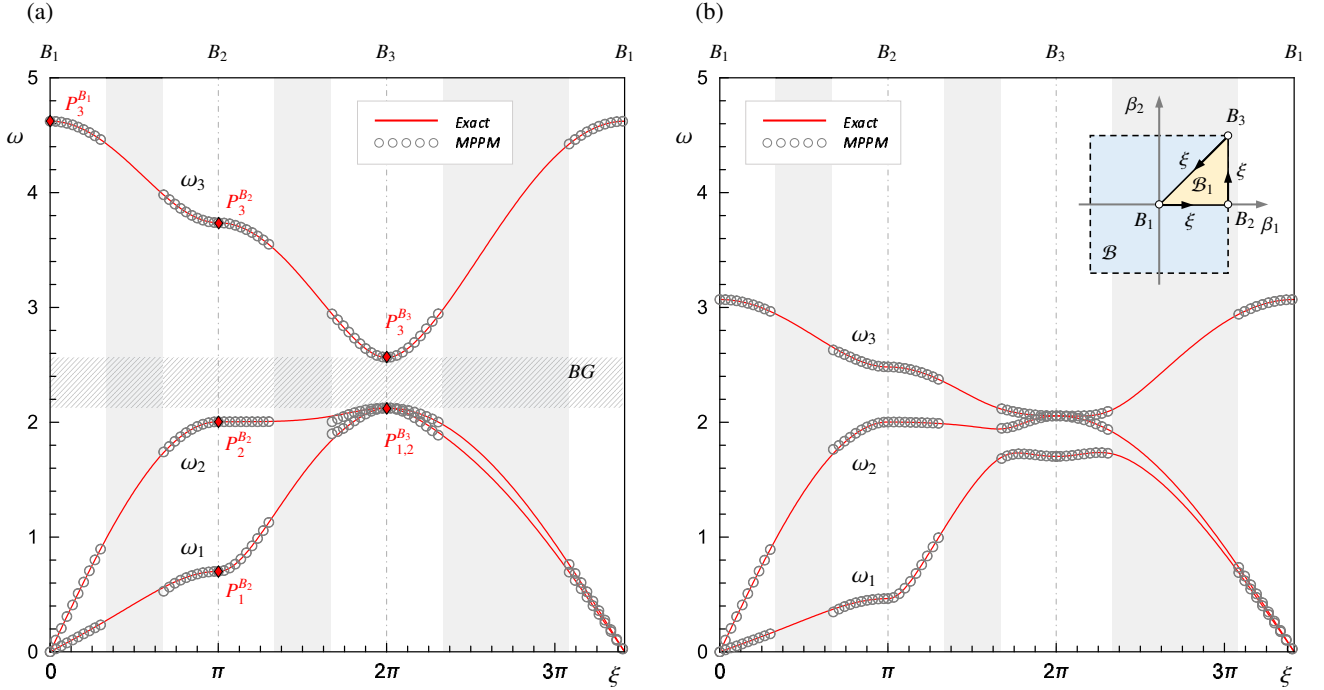

Figure 2: Floquet-Bloch spectrum of the tetrachiral material. Comparison between the exact and MPPM-approximate dispersion curves for (a) high-density, stiff material $(\delta, \varrho, \chi) = (1/10, 1/10, 1/9)$; (b) low-density, flexible material $(\delta, \varrho, \chi) = (1/15, 1/15, 1/9)$.

## 4. Tetrachiral material

The tetrachiral material is described by the mechanical parameters $\mathbf{p} = (\varrho, \delta, \chi)$. The Floquet-Bloch spectrum is characterized by three dispersion functions $\omega(\mathbf{p}, \mathbf{b})$. Figure 2 shows the dispersion curves along the entire boundary of the triangular $\mathcal{B}_1$-zone (spanned counterclockwise by the curvilinear abscissa $\xi \in [0, (2+\sqrt{2})\pi]$), for two different $\mathbf{p}$-values. A standard numerical solver has been used to determine the exact curves (continuous red lines). As interesting remark, Figure 2a shows that the material spectrum can exhibit a full band gap $BG$ between the low-frequency acoustic branches (curves $\omega_{1,2}$) and the high-frequency optical branch (curve $\omega_3$). Analyzing the polarization modes allows the clear recognition of a shear wave (eigenvector $\boldsymbol{\psi}_1$, corresponding to the frequency $\omega_1$), a compression wave ($\boldsymbol{\psi}_2$) and a rotational wave ($\boldsymbol{\psi}_3$) in the proximity of the point $\mathbf{b} = \mathbf{0}$ (limit case of long waves).

In order to apply the multi-parameter perturbation method, the mechanical parameters have been ordered according to the $\epsilon$-dependent law

$$\varrho = \epsilon \varrho', \qquad \delta = \epsilon \delta', \qquad \chi = \epsilon \chi' \tag{16}$$

whereas the wavenumbers obey to the law

$$\beta_1 = \beta_1^\circ + \epsilon \beta_1', \qquad \beta_2 = \beta_2^\circ + \epsilon \beta_2' \tag{17}$$

which is equivalent to assign $\mathbf{p}^\circ = (0,0,0)$, $\mathbf{p}' = (\varrho', \delta', \chi')$, $\mathbf{b}^\circ = (\beta_1^\circ, \beta_2^\circ)$, $\mathbf{b}' = (\beta_1', \beta_2')$.

The $\epsilon$-order smallness attributed to all the mechanical parameters is driven by physical considerations, specifically related to the microstructured chiral materials under investigation. Indeed, the ordering (16) is intended to account for the typical low-density distribution of the rings (small $\delta$), which usually possess a limited rotational-to-translational mass ratio (small $\chi$) and are interconnected by highly-slender ligaments (small $\varrho$).

The vertices $B_1, B_2, B_3$ of the triangular $\mathcal{B}_1$-zone have been selected as reference points in the $\mathcal{B}$-space. Therefore, three different sensitivity analyses have been performed, one for each of the corresponding $\mathbf{b}^\circ$-values (namely $\mathbf{b}_1^\circ, \mathbf{b}_2^\circ, \mathbf{b}_3^\circ$). The sensitivities $\lambda_i^{(n)}$ have been determined for all the eigenvalues (that is for $i = 1, 2, 3$) up the fourth order ($n = 4$). It may be worth remarking that the generating eigensolution is characterized by multiple eigenvalues ($m^\circ = 2$) for the vertices $B_1$ and $B_3$. The consequent indeterminacies at the lowest orders have required the solution of the perturbation equations up to the sixth-order to assess the fourth order eigensensitivities. The approximate dispersion functions are reported in the AppendixB (after reabsorption of the $\epsilon$-parameter and up to the second approximation order for the sake of shortness).

The approximate dispersion functions are marked by the black circles in Figure 2. For the material microstructures described by the particular $\mathbf{p}$-values, the fourth-order approximation shows a satisfying accuracy over large $\mathbf{b}$-ranges. To quantify, a cautious estimate of the approximation accuracy range for the wavevector perturbation is $|\mathbf{b}'| \leq 1$ (white windows centered in the vertices $B_1, B_2, B_3$). The amplitude $|\boldsymbol{\mu}'|$ of the full multi-parameter perturbation is certainly larger. By virtue of the specific treatment reserved to multiple eigenvalues, the asymptotic approximation well-performs also in the spec-tral regions with internal resonance (coincident frequencies) or nearly-resonance (close frequencies), where a stronger parameter sensitivity is expected. Of course, better approximation accuracies could be achieved by extending the approximation to higher orders or – in alternative – by adding and properly matching new local solutions starting from additional $\mathbf{b}^\circ$-points along the $\mathcal{B}_1$-boundary.

Table 1: Fourth-order approximation of the eigenvalues at the vertices of the $\mathcal{B}_1$-zone.

| Vertex | $B_1$ ($\beta_1=0, \beta_2=0$) | $B_2$ ($\beta_1=\pi, \beta_2=0$) | $B_3$ ($\beta_1=\pi, \beta_2=\pi$) |
|---|---|---|---|
| $\lambda_1$ | 0 | $24\varrho^2(2+5\delta^2)$ | $4+24\varrho^2(2+5\delta^2)+\tfrac{1}{2}\delta^2(4+3\delta^2)$ |
| $\lambda_2$ | 0 | $4+\tfrac{1}{2}\delta^2(4+3\delta^2)$ | $4+24\varrho^2(2+5\delta^2)+\tfrac{1}{2}\delta^2(4+3\delta^2)$ |
| $\lambda_3$ | $\frac{\delta^2}{4\chi^2}(8+\delta^2(4+3\delta^2))+\frac{3\varrho^2}{\chi^2}(8+3\delta^2(4+5\delta^2))$ | $\frac{\delta^2}{8\chi^2}(8+\delta^2(4+3\delta^2))+\frac{2\varrho^2}{\chi^2}(8+3\delta^2(4+5\delta^2))$ | $\frac{\varrho^2}{\chi^2}(8+3\delta^2(4+5\delta^2))$ |

### 4.1. Pass and stop band

The typical band structure of the tetrachiral material has been illustrated for two distinct parameter sets **p**, corresponding to different microstructures. Depending on the nondimensional ring diameter and ligament slenderness (parameters $\delta$ and $\varrho$, respectively), the first set physically identifies a high-density, stiff material (higher $(\delta, \varrho)$-values in Figure 2a). The second set corresponds instead to a low-density, flexible material (lower $(\delta, \varrho)$-values in Figure 2b). A rapid comparison indicates that reducing the material density and stiffness tends to close the full band gap $BG$ separating the low-frequency acoustic branches from the high-frequency optical branch. This systematic trend has been also confirmed by wider parametric analyses carried out all over the $\mathcal{B}$-zone by means of numerical optimization strategies (in analogy with [23]).

Moving from these qualitative observations, a challenging low-level task is to provide the parametric conditions for the existence of full band-gaps. Therefore, a higher-level task is to assess the amplitude of a certain band gap, if it exists, as an analytic function of the mechanical parameters. The analytic description of all the pass and stop bands may represent a matter of wider interest. For instance, the parametric tuning of the material spectrum could pave the way for the custom design of mechanical filters for elastic waves.

The multi-parameter perturbation solutions can be exploited to tackle these tasks, provided that all the hypotheses of parameter smallness are respected (that is, if $\delta, \varrho, \chi$ do not exceed $O(\epsilon)$, as postulated by the ordering (16)). Only the vertices $B_1, B_2, B_3$ of the $\mathcal{B}_1$-zone are assumed as *check points* to determine the limits of the pass and stop bands. If necessary, a larger number of check points, located either in $\mathcal{B}_1$-zone or on the $\partial\mathcal{B}_1$-boundary, could be adopted. Although this improvement would certainly strengthen the robustness and reliability of the analyses, it does not conceptually enrich the proposed methodology. The fourth-order asymptotic approximations of the eigenvalues $\lambda_i^{B_j}$ ($i=1,2,3$) at each vertex $B_j$ ($j=1,2,3$) are reported in Table 1. The red diamonds in Figure 2a mark the corresponding points $P_i^{B_j}$ in the $(\xi, \omega)$-space.

Recalling that all the mechanical parameters admit only strictly positive and small values (with smallness coherent with the ordering (16)), the eigenvalue approximations in Table 1 allow to draw the following remarks

- $\lambda_{1,2}^{B_1}$ and $\lambda_{1,2}^{B_3}$ are double eigenvalues in the whole **p**-domain. Indeed, they can split into couples of close eigenvalues only when $\mathbf{b}'$-perturbations are introduced to span the $\mathcal{B}_1$-region or the $\partial\mathcal{B}_1$-boundary (as in Figure 2).

- $\lambda_2^{B_2} > \lambda_1^{B_2}$, since they belong to different $\epsilon$-orders, namely $\lambda_2^{B_2} = O(\epsilon^0)$ while $\lambda_1^{B_2} = O(\epsilon^2)$. Therefore, considering also the previous remark, the first acoustic branch does not exceed the second acoustic branch (in the check points).

- $\lambda_2^{B_1} < \lambda_2^{B_2} < \lambda_2^{B_3}$, since $\lambda_2^{B_1} = 0$ and $\lambda_2^{B_3} - \lambda_2^{B_2}$ is a positive quantity. Consequently, the maximum of the second acoustic branch (restrained to the check points) is located at the vertex $B_3$, even if the difference $\lambda_2^{B_3} - \lambda_2^{B_2}$ is small (namely $O(\epsilon^2)$)

- $\lambda_3^{B_1} > \lambda_3^{B_2} > \lambda_3^{B_3}$, since $\lambda_2^{B_1} - \lambda_2^{B_3}$ is a positive quantity and also $\lambda_3^{B_2} = \tfrac{1}{2}(\lambda_3^{B_1}+\lambda_3^{B_3})$. Consequently, the minimum of the optical branch (restrained to the check points) is located at the vertex $B_3$.

As minor remarks, $\lambda_3^{B_1}$ identifies the upper bound of the band structure, according to the beam lattice model adopted for the tetrachiral material.

Recalling the relation $\lambda = \omega^2$, the above remarks highlight that the material spectrum can be essentially characterized by three frequency bands

- the *pass* band ranging from the lower (null) bound $\omega_1^{B_1}$ to the upper bound $\omega_2^{B_3}$

- the *stop* band ranging from the lower bound $\omega_2^{B_3}$ to the upper bound $\omega_3^{B_3}$

- the *pass* band ranging from the lower bound $\omega_3^{B_3}$ to the upper bound $\omega_3^{B_1}$

where the stop band actually exists only if $\omega_3^{B_3} > \omega_2^{B_3}$.

As interesting result following from the above remarks, the existence condition of the stop band can be expressed as a simple mathematical inequality among the mechanical parameters

$$\frac{\chi^2}{2\varrho^2} < \frac{8+3\delta^2(4+5\delta^2)}{8+\delta^2(4+3\delta^2)+48\varrho^2(2+5\delta^2)} \qquad (18)$$

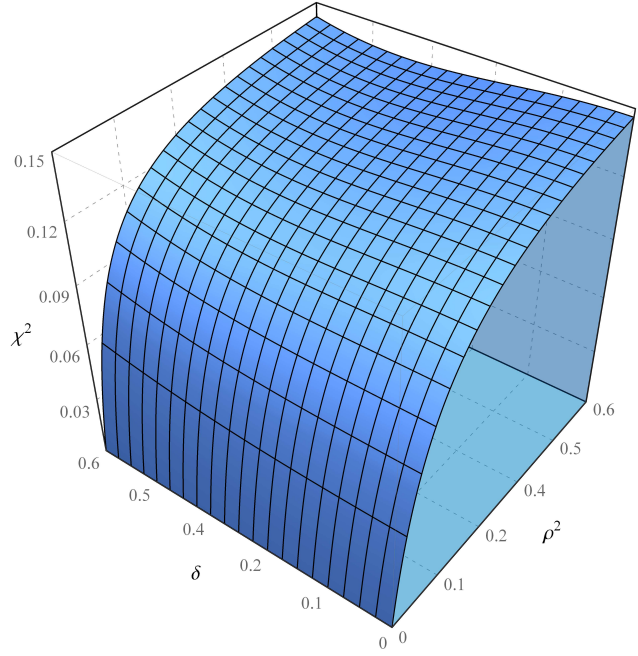

Figure 3: Parameter region $\mathcal{A}$ collecting all the **p**-points satisfying the existence condition for the stop band.

which is fully consistent with the fourth-order asymptotic approximations of the eigenvalues. From the physical viewpoint, the mathematical condition requires the rotational inertia of the ring to be a sufficiently small fraction of its mass (in dimensionless terms). The required smallness slightly depends on the elastic and geometric properties of the cellular microstucture.

The parameter region $\mathcal{A}$, which collects all the **p**-points satisfying the condition (18), is shown in Figure 3 (blue filled volume). From a qualitative perspective, it can be noticed that the $\mathcal{A}$-boundary is minimally affected by the $\delta$-parameter, while it strongly depends on the $\varrho$-parameter. To specify, increasing the ligament slenderness (lower $\varrho$-values) requires smaller $\chi$-values to keep open the band gap. From a quantitative standpoint, it is worth noting that reducing the ring mass ratio $\chi$ below a certain smallness tends to conflict with geometric constraints. Nonetheless, these geometric constraints can be relaxed by adopting suited constructional solutions, such as filling the ring with heavy material or using tapered cross-section.

Within the $\mathcal{A}$-region, the multi-parametric perturbation method provides also a fourth-order analytical approximation for the band gap amplitude

$$\Delta_s \simeq \Delta_s^\circ + \Delta_s'' + \Delta_s'''' \tag{19}$$

where the number of apices allows to recognize the $\epsilon$-order of each contribution in the sum (the absence of odd-order terms can be noted). The terms of the sum read

$$\Delta_s^\circ = 2\,(\varrho_s - 1) \tag{20}$$

$$\Delta_s'' = \tfrac{1}{2}\,\delta^2\,(3\varrho_s - 1) - 12\varrho^2 \tag{21}$$

$$\Delta_s'''' = \tfrac{1}{16}\,\delta^4\,(21\varrho_s - 5) + 9\varrho^2(4\varrho^2 - 3\delta^2) \tag{22}$$

where the $O(\epsilon^0)$-order combination parameter $\varrho_s^2 = 2\varrho^2/\chi^2$ plays a key role. Indeed, it can be verified that the condition (18) is asymptotically equivalent to

$$\varrho_s^2 > 1 + \left(12\varrho^2 - \delta^2\right) + 12\varrho^2\delta^2 \tag{23}$$

which essentially requires $\varrho_s^2 > 1$, if higher orders are neglected. This low-order condition is fully consistent with the asymptotic approximation (19) obtained for the band gap amplitude. Indeed, if this low-order condition for the existence of the band gap is satisfied, the lowest-order contribution to the band gap amplitude certainly attains a positive value ($\Delta_s^\circ > 0$).

The band gap amplitude $\Delta_s$ is illustrated in Figure 4 for different sections of the $\mathcal{A}$-region, obtained by fixing one of the three parameters in the formula (18). Gray zones are outside of the $\mathcal{A}$-region and do not admit the existence of band gap.

Figure 4 shows that the lowest-order asymptotic approximation $\Delta_s^\circ$ can be sufficient to well-describe the essential parametric dependence of the band gap amplitude. Indeed, the amplitude strongly varies with the two important parameters $\varrho^2$ (with almost linear *direct* $\varrho$-proportionality) and $\chi^2$ (with almost linear *inverse* $\chi$-proportionality). This remark fits with the $\varrho_s$-dependence of the dominant, zeroth-order term $\Delta_s^\circ$ in the equation (20). On the contrary, the amplitude weakly depends on the third parameter $\delta$, consistently with the $\delta$-dependence of the second-order term $\Delta_s''$ in the equation (21). From the physical viewpoint, these results state that larger band gaps can be obtained either by reducing the ligament slenderness (higher $\varrho$-values), or by reducing the rotational inertia of the rings (lower $\chi$-values). Differently, the band gap amplitude is less sensitive to variations of the ring diameter (different $\delta$-values).

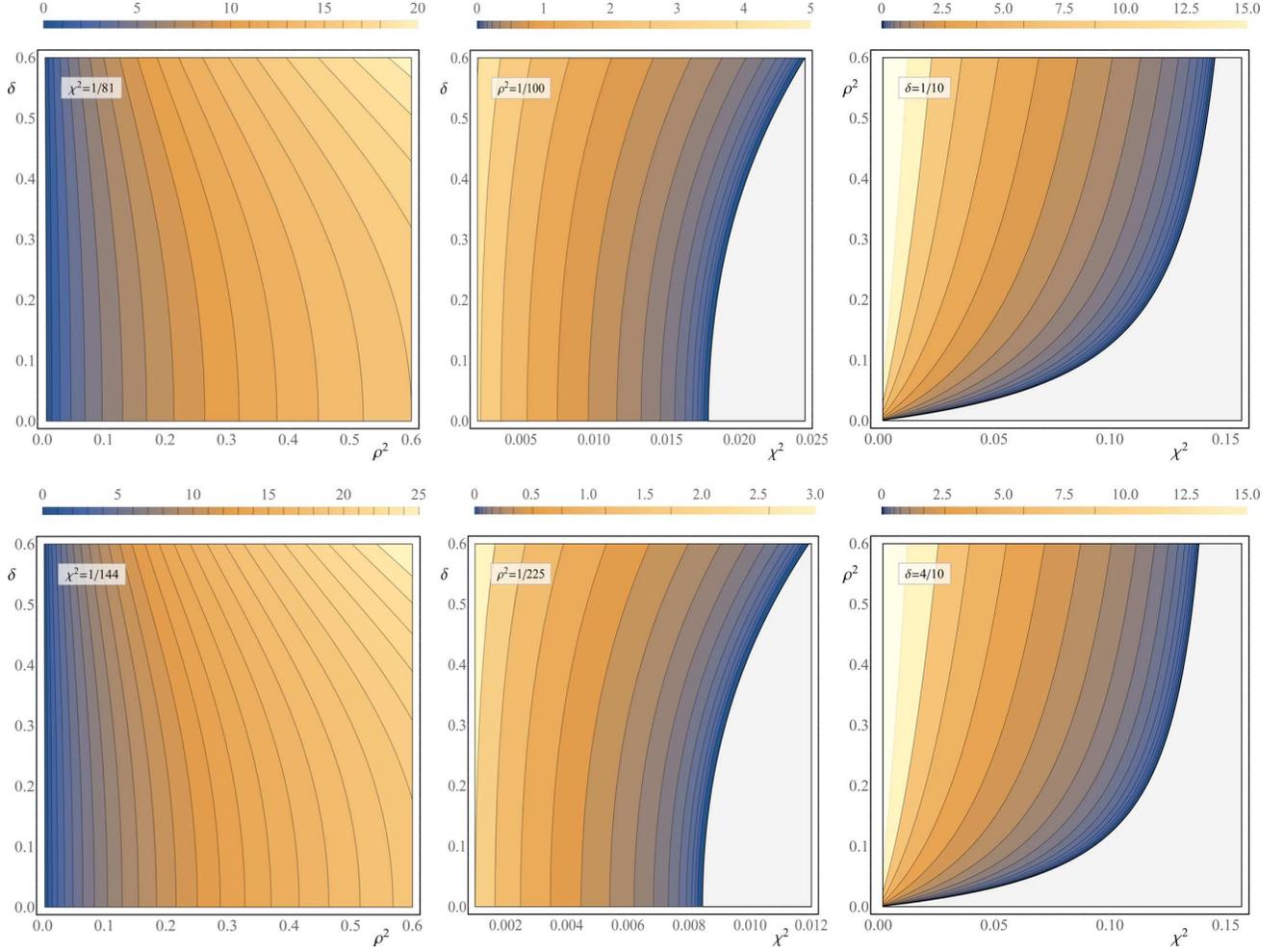

Figure 4: Band gap amplitude versus the mechanical parameters $\mathbf{p} = (\varrho, \delta, \chi)$. Note that gray regions are outside of the $\mathcal{A}$-region, while null values of the $(\varrho, \chi)$-parameters are physically inadmissible.

## 5. Tetrachiral metamaterial

Due to the introduction of the inertial resonator, the metamaterial microstructure is described by a larger set of mechanical parameters $\mathbf{p} = (\varrho, \delta, \chi, \gamma, \alpha, \gamma_\theta, \chi_\mathbf{r})$. Moreover, the increment of the active degrees-of-freedom enriches the Floquet-Bloch spectrum, which is featured by six dispersion functions $\omega(\mathbf{p}, \mathbf{b})$.

Figure 5a shows the metamaterial spectrum along the boundary of the $\mathcal{B}_1$-zone for selected values of the $\mathbf{p}$-parameters. The six dispersion curves (blue lines) can be compared with the three curves of the corresponding tetrachiral material (red dashed lines). The comparison shows how the new spectrum branches generated by the inertial resonators significantly modify the material band structure. The principal effect is the opening of the new full band-gaps $BG_1$ and $BG_2$, which are approximately centered at the two frequencies pointed by the tuning parameters $\gamma$ and $\gamma_\theta$, respectively. The band-gaps $BG_1$, in particular, is centered in the low-frequency range and its amplitude is directly dependent on the resonator mass $\alpha$.

The multi-parameter perturbation method can straightforwardly be extended to tackle the eigenproblem governing the metamaterial spectrum. Indeed, the mathematical formulation presented in Section 3.1 does not require any formal adjustment to deal with the higher dimension of the configuration vector $\mathbf{q}$ and the enlarged parameter set $\mathbf{p}$. However, to simplify the discussion, the parameter set can suitably be decomposed in the form $\mathbf{p} = (\mathbf{p}_s, \mathbf{p}_r)$, where the subset $\mathbf{p}_s = (\varrho, \delta, \chi)$, collecting the mechanical parameters of the cell microstructure, is distinguished from the subset $\mathbf{p}_r = (\gamma, \alpha, \gamma_\theta, \chi_r)$, collecting the mechanical parameters of the resonator.

Recalling that the ordering (16) has been already adopted for the microstructural parameters $\mathbf{p}_s$, the following ordering can be introduced for the resonator parameters $\mathbf{p}_r$

$$\alpha = \epsilon \alpha', \quad \gamma = \gamma^\circ + \epsilon \sigma, \quad \chi_r = \epsilon \chi'_r, \quad \gamma_\theta = \gamma_\theta^\circ \qquad (24)$$

which is equivalent to assign the reference parameter set $\mathbf{p}^\circ = (0, 0, 0, \gamma^\circ, 0, \gamma_\theta^\circ)$ and the multi-parameter perturbation $\mathbf{p}' = (\varrho', \delta', \chi', \alpha', \sigma, \chi'_r, 0)$. The additional condition $\gamma_\theta^\circ - \gamma^\circ = O(\epsilon^0)$ is assumed for the sake of simplicity.

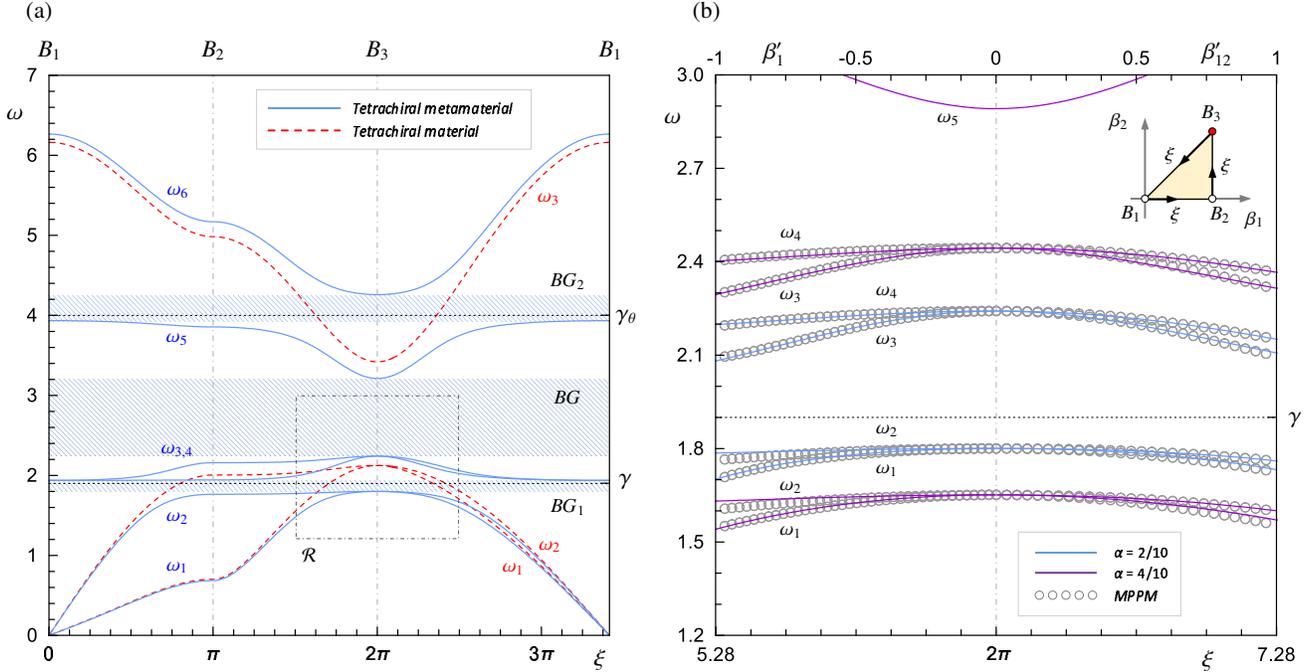

Figure 5: Floquet-Bloch spectrum of the tetrachiral metamaterials with microstructural parameters $(\delta, \varrho, \chi) = (1/10, 1/10, 1/12)$ and resonator parameters $(\alpha, \gamma, \chi_r, \gamma_\theta) = (2/10, 19/20, 1/11, 4)$: (a) comparison with the spectrum of the tetrachiral material, (b) local comparison of the exact versus MPPM-approximate dispersion curves in the $\mathcal{R}$-region for different resonator masses.

From the physical viewpoint, the smallness of the parameter $\alpha$ is intended to account for *light* resonators, whose mass is not greater than that of the hosting ring. The ordering of the parameter $\gamma$ replicates instead the typical spectral design of the inertial resonator, which is primarily tuned to the target frequency $\gamma^\circ$, with $\sigma$ playing the role of *internal detuning*.

From a mechanical perspective, it may be worth recalling that the resonator stiffness is a dependent parameter. In particular, the parameter ordering (24) implicitly requires that the resonator stiffness has the same $\epsilon$-order of the resonator mass. Otherwise, their ratio – i.e. the resonator frequency – could not remain finite-valued when the resonator mass tends to vanish (or, mathematically, $\gamma$ could not be $O(\epsilon^0)$ for $\epsilon \to 0$).

In order to severely check the effectiveness of the multi-parameter perturbation method, a parameter region corresponding to a high spectral density is preferable. To this aim, a local sensitivity analysis has been focused on the $\mathcal{R}$-zone of the material spectrum, which is centered at the $B_3$-vertex and featured by a cluster of four dispersion curves (Figure 5a).

Therefore, the reference points $\mathbf{p}^\circ$ and $\mathbf{b}^\circ$ have been fully assigned by setting $\gamma^\circ = 2$, $\gamma_\theta^\circ = 4$ and fixing $\beta_1^\circ = \beta_2^\circ = \pi$, respectively. Physically, this selection of the zeroth-order parameters corresponds to tuning the inertial resonator with the double frequency $\omega_{1,2}^{B_3}$ of the tetrachiral material spectrum (see Figure 2a and Table 1). Mathematically, in the tuned metamaterial the solution algorithm for the perturbation equations is complicated by the occurrence of a high eigenvalue multiplicity ($m^\circ = 4$) in the zeroth-order solution. To specify, the eighth-order of the perturbation equations must be solved to clear all the indeterminacies and carry out the fourth-order eigensensitivities.

Figure 5b shows the exact dispersion curves of the metamaterial in the $\mathcal{R}$-zone for two different values of the resonator mass. The corresponding fourth-order approximation is marked by the black circles. A satisfying agreement between the exact and approximate curves can be appreciated over a large $\mathbf{b}'$-range (say for $2\pi - 1 \leq \xi \leq 2\pi + 1$, corresponding to $|\mathbf{b}'| \leq 1$, or even more). Therefore, it can be remarked that neither the enlargement of the multi-parameter perturbation vector $\mathbf{p}'$ nor the higher density of the metamaterial spectrum compromise the performance of the asymptotic technique.

Fixing the wavevector and the cell microstructure, the multi-parametric approximate solution can be exploited to explore the metamaterial band structure in the $(\alpha, \sigma)$-subspace of the resonator parameters. The fourth-order approximation of the double eigenvalues $\lambda_{1,2}$ and $\lambda_{3,4}$ in the $B_3$-vertex is reported in the AppendixB.5. Figure 6 shows the corresponding frequency loci $\omega_{1,2}$ and $\omega_{3,4}$ in the plane spanned by the $(\alpha, \sigma)$-parameters. The asymptotically approximate loci (blue surfaces in Figure 6b) are found to well-match the exact loci (yellow surfaces in Figure 6a) over the full range of investigation, bounded by small $(\alpha, \sigma)$-values for the sake of consistency.

It is worth remarking that the asymptotic solution may be not uniformly valid in the whole $(\alpha, \sigma)$-plane. Indeed, the asymptotic approximation turns out to break down for very small mass ratios, corresponding to *extra-light* resonators (approximately for $\alpha^2 \leq 4 \times 10^{-4}$). The validity loss is determined by vertical asymptotes in the approximation function and reflects in the spurious narrow peak featuring the approximate frequency loci, close to the origin of the $(\alpha^2, \sigma)$-plane (Figure 6b). This mathematical singularity is not a fault in the multi-parameter perturbation approach. Actually, it can be regarded as a sort of boundary layer issue, inherent to the particular parameter ordering postulated in the equation (24) to build up the asymptotic solution (say the *outer* solution). Consequently, the parameter ordering must be partially modified (by setting $\alpha = \epsilon^2 \alpha'$ and $\gamma = \gamma^\circ + \epsilon^2 \sigma$), in order to determine a distinct asymptotic solution (*inner* solution), suited to well approximate the exact solution within the boundary layer. This specific issue has been examined in more depth in [41], whereas some additional notes on the matter are reported in the AppendixB.5. Here it can be sufficient to highlight how the inner solution is valid in the neighborhood of the origin, although its approximation is less accurate than the outer solution in the rest of the $(\alpha^2, \sigma)$-plane (green surfaces in Figure 6b). As common in perturbation methods, a proper matching of the two asymptotic solutions could combine the validity of the inner solution with the approximation accuracy of the outer solution, if necessary.

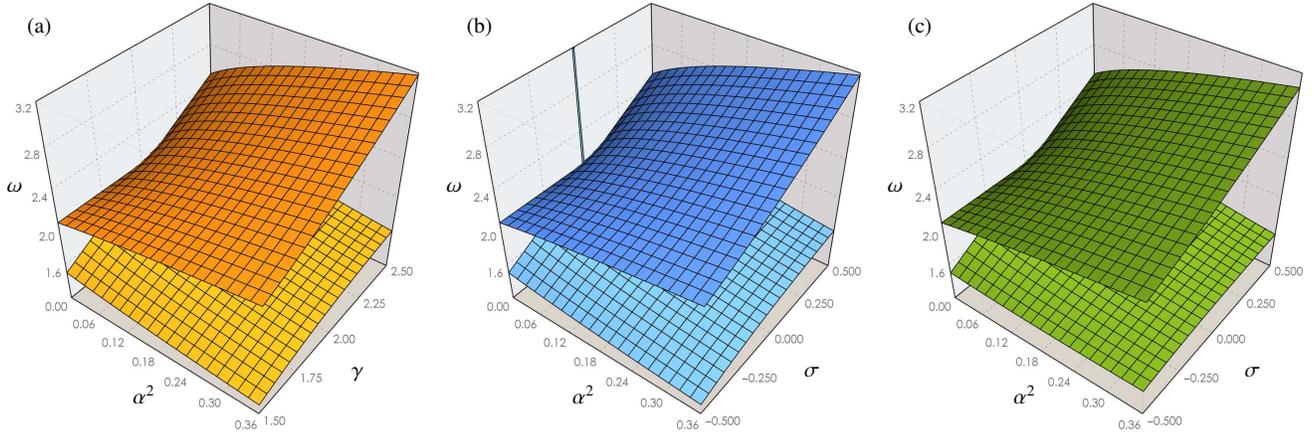

Figure 6: Frequency loci vs the resonator parameters for the tetrachiral metamaterial with $(\delta, \varrho, \chi) = (1/10, 1/15, 1/12)$: (a) exact, (b) MPPM-approximation for light resonators, (c) MPPM-approximation for extra-light resonators.

### 5.1. Stop band amplitude

Estimating the stop bandwidths in the low-frequency range is a key point to analyze the potential of periodic metamaterials as mechanical filters. According to the multiparameter perturbation method, the problem can be synthesized as follows.

> For a certain cellular microstructure fixed by the parameters $\mathbf{p}_s$, determine how the amplitude $\Delta_r$ of the stop band located across the resonator frequency $\gamma^\circ$ depends on the resonator mass $\alpha^2$ and detuning $\sigma$.

Focusing on the band structure of the tetrachiral metamaterial, the lowest frequency stop band located across the resonator frequency $\gamma^\circ = 2$ (band gap $BG_1$ in Figure 5) is considered. The band gap lies between the second and third dispersion curves. In the general case the band gap is bounded by the frequencies $\omega_2^{B_3}$ (lower bound) and $\omega_3^{B_1}$ (upper bound). Therefore, the vertices $B_1$ and $B_3$ of the $\mathcal{B}_1$-zone are assumed as check points to first verify the existence of the band gap, and then assess its amplitude. The fourth-order asymptotic approximations of the eigenvalues $\lambda_3^{B_1}$ and $\lambda_2^{B_3}$ are reported in the AppendixB.5.

According to the selection of the check points, the band gap amplitude $\Delta_r$ admits the fourth-order asymptotic approximation

$$\Delta_r \simeq \Delta_r' + \Delta_r'' + \Delta_r''' + \Delta_r'''' \qquad (25)$$

where the number of apices allows to recognize the $\epsilon$-order of each contribution in the sum. The $n$-th order term $\Delta_r^{(n)}$ reads

$$\Delta_r' = \tfrac{1}{2}(\sigma + S_1) \qquad (26)$$

$$\Delta_r'' = \tfrac{1}{4}\left(3\alpha^2 - d_2^2\right) + \tfrac{1}{4}\sigma\left(4\alpha^2 - d_2^2\right) S_1^{-1} \qquad (27)$$

$$\Delta_r''' = \tfrac{5}{16}(\alpha^2 \sigma) + \tfrac{1}{4}\alpha^2 \left(\sigma^2 + d_2^2\right)^2 S_1^{-3} \qquad (28)$$

$$\Delta_r'''' = \tfrac{1}{64}\left(\alpha^2\left(7\sigma^2 - 15\alpha^2\right) + 2d_2^2\left(\alpha^2 + d_2^2\right)\right) + \qquad (29)$$
$$\quad - \tfrac{1}{64}\alpha^2\sigma\left(16\alpha^2\left(5\alpha^2 + 2\sigma^2\right) + 7\sigma^4\right)S_1^{-3} +$$
$$\quad + \tfrac{1}{32}\sigma d_2^2\left(d_2^2\left(2\alpha^2 + \sigma^2\right) - \alpha^2\left(12\alpha^2 + 7\sigma^2\right)\right)S_1^{-3} +$$
$$\quad - \tfrac{1}{4}\sum_{i=0}^{6}\frac{\alpha^{2i}}{D_1}\Big(\sum_{j=0}^{5}c_{ij}\sigma^{2j} + S_1\sum_{j=1}^{5}c_{ij}^*\sigma^{2j-1}\Big)$$

where $S_1 = (4\alpha^2 + \sigma^2)^{1/2}$ is a characteristic quantity of the resonator, while the auxiliary coefficients $c_{ij}, c_{ij}^*, d_2$ depend only on the mechanical parameters of the cell microstructure (see the AppendixB.5). Finally, the denominator $D_1$ depends on both the microstructural and the resonator parameters.

Within the physical limits established by the smallness hypotheses postulated for the parameters, the asymptotic approximation (25) allows to disclose some characteristic properties of the microstructure-resonator interaction governing the band gap amplitude

- the amplitude $\Delta_r$ does not possess any term belonging to $O(\epsilon^0)$. Recalling that $\gamma^\circ$ is $O(\epsilon^0)$, this asymptotic result means that light resonators cannot physically realize a frequency stop bandwidth larger than their own frequency

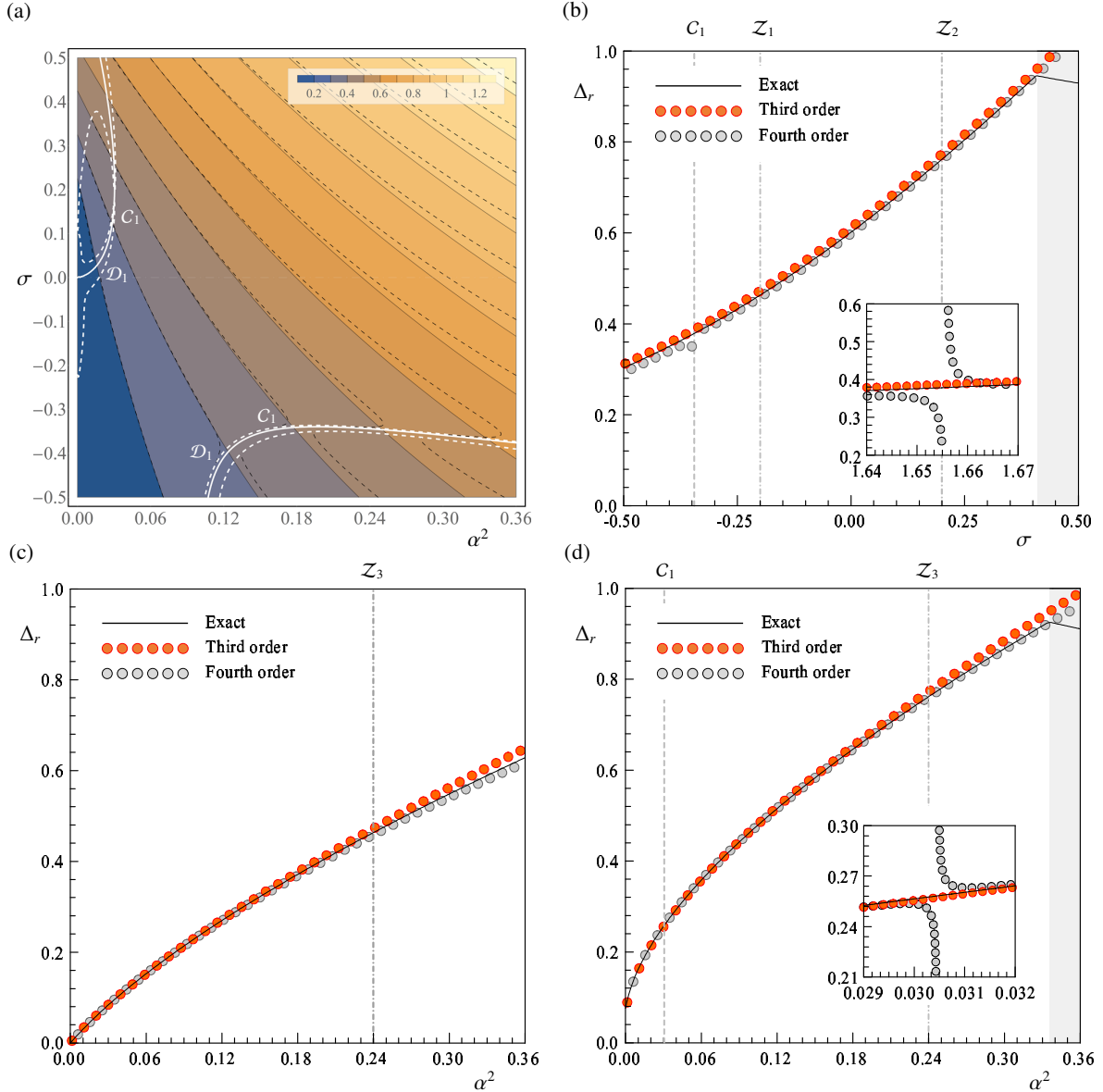

Figure 7: Amplitude of the band gap $BG_1$ in the $(\alpha^2, \sigma)$-space of the resonator parameters for fixed $(\delta, \varrho, \chi) = (1/10, 1/10, 1/12)$ and $(\gamma_\theta, \chi_r) = (4, 1/10)$: (a) third-order (continuous lines) vs fourth-order approximation (dashed lines). Comparison of the exact vs approximate amplitude in (b) section $\mathcal{Z}_3$ for $\alpha^2 = 24/100$, (c) section $\mathcal{Z}_1$ for $\sigma = -1/10$, (d) section $\mathcal{Z}_2$ for $\sigma = 1/10$.

- the dominant (lowest-order) term $\Delta'_r$ of the band gap amplitude is certainly positive for strictly positive $\alpha^2$-values. This mathematical result states that the inertial resonator is always able to open a small-amplitude band gap, independently of its lightness

- the dominant (lowest-order) term $\Delta'_r$ of the band gap amplitude depends only on the resonator parameters. This asymptotic result states that the essential contribution to the stop bandwidth is practically independent of the mechanical parameters of the cell microstructure.

The last remark, in particular, strongly indicates that the cellular microstructure and the light resonators may possess a quasi-independent dynamics. Indeed, the asymptotic eigensolutions reported in the AppendixB.5 reveal how the eigenvalues $\lambda_2^{B_3}$ and $\lambda_3^{B_1}$ are $\mathbf{p}_s$-independent up to $O(\epsilon)$ and correspond to waveforms strongly localized in the resonators. It can be remarked how a similar behavior has been recurrently highlighted in the perturbation-based eigensensitivity analysis of different structural systems equipped by small resonant masses, like tuned mass dampers [42, 43] or light substructural elements [44, 45].

Figure 7 shows the amplitude $\Delta_r$ of the band gap $BG_1$ in the $(\alpha^2, \sigma)$-space of the resonator parameters for a fixed cellular microstructure. The effects of an increasing mass $\alpha^2$ for both *undertuned* (detuning $\sigma < 0$) and *overtuned* ($\sigma > 0$) resonators can be appreciated in Figure 7a. On the one hand, overtuned res-onators open larger band gaps than undertuned resonators with the same mass. This systematic trend can be clearly observed in the $(\Delta_r, \sigma)$-section plane $\mathcal{Z}_3$ in Figure 7b, where the amplitude monotonically grows up for increasing detunings, fixed a certain $\alpha^2$-value. On the other hand, larger masses lead to higher amplitudes, no matter the resonator detuning. This behaviour is shown in the $(\Delta_r, \alpha^2)$-section planes $\mathcal{Z}_1$ and $\mathcal{Z}_2$, where the amplitude increment versus the mass is illustrated for an undertuned and an overtuned resonators (Figures 7c,d). The highest amplitudes can be achieved for highly massive and strongly overtuned resonators. However, is must be noted that amplitude values greater than unity tend to lie in the exclusion region (gray zones in Figures 7b,d), in which the check points do not correspond to the limits of the stop band.

Figure 7 shows also the minor differences between the third-order and the fourth-order approximation of the band gap amplitude. As general remark, the third-order approximation tends to slightly overestimate the amplitude. Nonetheless, the quantitative comparison between the exact and the approximate amplitudes shows a satisfying agreement up to large parameter values, apart the exclusion regions (Figures 7b,c,d).

As minor remark, the two curves $C_1$ in Figure 7a describe the $(\alpha^2, \sigma)$-loci in which the fourth order approximation breaks down. Indeed, all the $C_1$-points determine a singularity in the parametric function expressing the fourth sensitivity of the eigenvalue $\lambda_2^{B_3}$. Consequently, the fourth order approximation can be observed to diverge over a narrow range of the varying parameters, centered around the $C_1$-points (see the windows in Figures 7b,d). The regions $\mathcal{D}_1$, bounded by the white dashed lines and enclosing the $C_1$-curves, collect the $(\alpha^2, \sigma)$-points in which the fourth order approximation is expected to fail, because the fourth sensitivity of the eigenvalue exceeds the order of the lower sensitivities. A deeper discussion about this issue is reported in the AppendixB.5.

### 5.2. Inertial resonator design

According to the multiparameter perturbation method, the inverse problem related to design a mechanical filter by equipping the tetrachiral material with inertial resonators could be synthesized as follows.

> For a certain cellular microstructure fixed by the parameters $\mathbf{p}_s$, determine the resonator parameters (frequency $\gamma^\circ$, mass $\alpha^2$ and internal detuning $\sigma$), which modify the material spectrum by opening a stop band characterized by the desired bandwidth $\tilde{\Delta}_r$ centered at the assigned center frequency $\tilde{\gamma}_c$.

The problem solution is not trivial because the stop bandwidth is a nonlinear function of the resonator parameters. Furthermore – in the absence of detuning – the center frequency of the stop band turns out to be close but not identical to the resonator frequency $\gamma^\circ$. Therefore, the detuning $\sigma$ must be intended as the free parameter required to compensate this tuning defect. The solution can be found by exploiting the perturbation method, under the assumption that the tuning defect is small enough (that is, if the available $\gamma^\circ$-range of the resonator frequencies satisfies the condition $\tilde{\gamma}_c - \gamma^\circ = O(\epsilon)$).

From the mathematical (asymptotic) viewpoint, the design problem is governed by a nonlinear system of algebraic equations. Selected the proper check points, the first equation imposes their frequency difference $\Delta_r$ to match the assigned stop bandwidth. The second equation forces the frequency half-sum $\frac{1}{2}\Sigma_r$ to be equal to the assigned center frequency. At the fourth order of approximation, the equations read

$$\Delta'_r + \Delta''_r + \Delta'''_r + \Delta''''_r = \tilde{\Delta}_r \qquad (30)$$
$$\gamma^\circ + \frac{1}{2}(\Sigma'_r + \Sigma''_r + \Sigma'''_r + \Sigma''''_r) = \tilde{\gamma}_c \qquad (31)$$

For the specific metamaterial under investigation (with $\gamma^\circ = 2$) the two equations allow to determine the unknown detuning $\sigma$ and $\alpha^2$. Of course, the solution exists if the data of the inverse problem are consistent with the solution of the direct problem (for instance, the assigned amplitude $\tilde{\Delta}_r$ cannot exceed $O(\epsilon)$).

The search for the $(\alpha^2, \sigma)$-solution is graphically represented in Figure 8. The curve $\tilde{\mathcal{D}}_r$ is the locus of $(\alpha^2, \sigma)$-points satisfying the equation (30). Actually, it can be recognized as one of the contour lines (selected according the assigned amplitude level $\tilde{\Delta}_r = 6/10$) of the amplitude contour map in Figure 7a. The curve $\tilde{\mathcal{G}}_1$ is instead the locus of $(\alpha^2, \sigma)$-points satisfying the equation (31) for a certain assigned center frequency ($\tilde{\gamma}_c = 19/10$). The unique intersection point identifies the solution. It can be noted that the solution point $\tilde{\mathcal{P}}'''_1$ returned by the third order equations is slightly different from the solution point $\tilde{\mathcal{P}}''''_1$ returned by the fourth order equations. In particular, the third order solution tends to slightly underestimate the parameter values. The same considerations can be extended to the crossing points $\tilde{\mathcal{P}}_2'''$ and $\tilde{\mathcal{P}}_2''''$ between the curves $\tilde{\mathcal{D}}_r$ and $\tilde{\mathcal{G}}_2$, which refers to a different assignment of the center frequency ($\tilde{\gamma}_c = 21/10$).

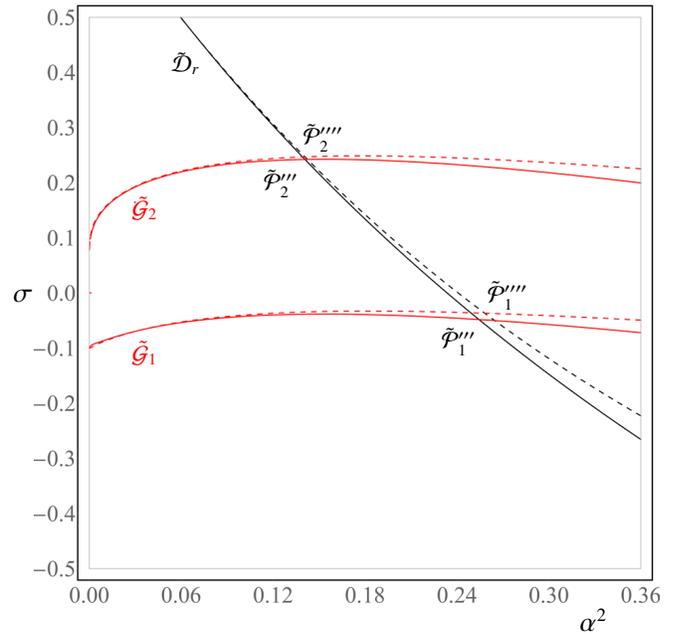

Figure 8: Graphical scheme for the design of the band gap $BG_1$ in the $(\alpha^2, \sigma)$-space of the resonator parameters for fixed $(\delta, \varrho, \chi) = (1/10, 1/10, 1/12)$ and $(\gamma_\theta, \chi_r) = (4, 1/10)$, based on the third-order (continuous lines) or the fourth-order approximation (dashed lines).

Table 2: Design parameters ($\alpha^2, \sigma$) of the resonator for given stop bandwidth $\tilde{\Delta}_r$ and center frequency $\tilde{\gamma}_c$.

| | | $\tilde{\gamma}_c = 1.9$ | | | $\tilde{\gamma}_c = 2$ | | | $\tilde{\gamma}_c = 2.1$ | | |
| --- | --- | --- | --- | --- | --- | --- | --- | --- | --- | --- |
| | | Third | Fourth | Check | Third | Fourth | Check | Third | Fourth | Check |
| $\tilde{\Delta}_r = 0.5$ | $\alpha^2$ | 0.1941 | 0.1958 | $\Delta_r = 0.5018$ | 0.1454 | 0.1456 | $\Delta_r = 0.5003$ | 0.1013 | 0.1011 | $\Delta_r = 0.4998$ |
| | $\sigma$ | −0.04017 | −0.03302 | $\gamma_c = 1.9000$ | 0.09755 | 0.1022 | $\gamma_c = 1.9999$ | 0.2367 | 0.2394 | $\gamma_c = 2.0999$ |
| $\tilde{\Delta}_r = 0.6$ | $\alpha^2$ | 0.2544 | 0.2581 | $\Delta_r = 0.6035$ | 0.1946 | 0.1955 | $\Delta_r = 0.6010$ | 0.1408 | 0.1407 | $\Delta_r = 0.5997$ |
| | $\sigma$ | −0.04822 | −0.03671 | $\gamma_c = 1.9003$ | 0.09608 | 0.1039 | $\gamma_c = 1.9999$ | 0.2422 | 0.2468 | $\gamma_c = 2.0998$ |
| $\tilde{\Delta}_r = 0.7$ | $\alpha^2$ | 0.3226 | 0.3296 | $\Delta_r = 0.7063$ | 0.2509 | 0.2529 | $\Delta_r = 0.7020$ | 0.1867 | 0.1869 | $\Delta_r = 0.7000$ |
| | $\sigma$ | −0.06255 | −0.04502 | $\gamma_c = 1.9011$ | 0.08808 | 0.1004 | $\gamma_c = 2.0000$ | 0.2408 | 0.2486 | $\gamma_c = 2.9998$ |
| $\tilde{\Delta}_r = 0.8$ | $\alpha^2$ | 0.3996 | 0.4116 | $\Delta_r = 0.8109$ | 0.3147 | 0.3187 | $\Delta_r = 0.8038$ | 0.2393 | 0.2401 | $\Delta_r = 0.8005$ |
| | $\sigma$ | −0.08303 | −0.05745 | $\gamma_c = 1.9024$ | 0.07366 | 0.0920 | $\gamma_c = 2.0004$ | 0.2329 | 0.2449 | $\gamma_c = 2.0997$ |

Table 2 collects the design problem solutions for a few case-studies related to different bandwidths $\tilde{\Delta}_r$. It must be remarked that greater masses are required to open stop bands with the same amplitude at lower center frequencies. In the particular case $\tilde{\gamma}_c = 2$, the desired center frequency $\tilde{\gamma}_c$ of the stop band is equal to the resonator frequency $\gamma°$. The solution highlights that a small overtuning ($\sigma > 0$) of the inertial resonator is always required to satisfy the design target. Otherwise, in the absence of detuning ($\sigma = 0$) the stop bandwidth would be located at a lower center frequency.

As complementary analytical achievement, the design problem can be solved parametrically, by adopting an asymptotic approach to the governing equations (30),(31). Although approximate, the parametric solution may be useful for the preliminary evaluation of the design parameters required to open the band gap characterized by the desired ($\tilde{\Delta}_r, \tilde{\gamma}_c$)-values. According to this approach, two design formulas can be given for the resonator mass ratio $\alpha^2$ and frequency $\gamma$

$$\alpha^2 = \tilde{\alpha}'' + \tilde{\alpha}''' + \tilde{\alpha}'''' \tag{32}$$

$$\gamma = \gamma° + \gamma' + \gamma'' + \gamma''' \tag{33}$$

where the terms of the sums are

$$\alpha'' = \tfrac{1}{2}\tilde{\Delta}_r(\tilde{\Delta}_r - 2\tilde{\Gamma}_c) \tag{34}$$

$$\alpha''' = \tfrac{1}{16}\tilde{\Delta}_r\left(\tilde{\Delta}_r(8\tilde{\Gamma}_c - 5\tilde{\Delta}_r) + 4(\tilde{\Gamma}_c^2 + 2\delta^2 + 48\varrho^2)\right) \tag{35}$$

$$\alpha'''' = \tfrac{1}{32}\tilde{\Delta}_r\left(\tilde{\Delta}_r(12\tilde{\Delta}_r - 33\tilde{\Gamma}_c) + 20\tilde{\Gamma}_c^2 - 16\delta^2)\right) + \tag{36}$$
$$+ \tfrac{1}{32}\tilde{\Delta}_r\left(4\tilde{\Gamma}_c(2\delta^2 - \tilde{\Gamma}_c^2) + 192(\tilde{\Gamma}_c - 2\tilde{\Delta}_r)\varrho^2\right)$$

$$\gamma' = \tfrac{1}{2}\tilde{\Delta}_r + \tilde{\Gamma}_c \tag{37}$$

$$\gamma'' = \tfrac{1}{2}\tilde{\Delta}_r\left(2\tilde{\Gamma}_c - \tilde{\Delta}_r\right) \tag{38}$$

$$\gamma''' = \tfrac{1}{16}\tilde{\Delta}_r\left((\tilde{\Delta}_r - 2\tilde{\Gamma}_c)(3\tilde{\Delta}_r - 2\tilde{\Gamma}_c) - 8(\delta^2 + 24\varrho^2)\right) \tag{39}$$

where $\tilde{\Gamma}_c = \tilde{\gamma}_c - \gamma°$ and $\gamma° = 2$ for the specific band gap to be opened in the metamaterial under investigation.

In summary, this perturbation-based methodology is sufficiently flexible to treat similar inverse spectral problems, connected to the pass and stop band design in periodic materials and metamaterials. On the other hand, it must be stressed that – as natural for local eigensensitivity analyses – the solutions are asymptotically approximate. Therefore, the results are inevitably case-specific, or strictly consistent with the smallness hypotheses assumed in the ordering of the governing parameters, or – finally – somehow related with the expert choice of the number and location of the starting points for the perturbation analysis (playing the role of check points for the band gap search). In this respect, it is understood that the asymptotic solutions of the design problem must be verified a posteriori, by checking the validity of the simplifying assumptions and the expert choices (for instance, through a complete investigation of the dispersion surfaces over the first Brillouin zone for the design parameters). In the case-study under investigation, the exact stop bandwidth and the exact center frequency corresponding to the design solution (based on the perturbation method) have been successfully checked a posteriori. The properties of the total band gap opened by the design solutions, checked over the first Brillouin zone, are reported in Table 2.

## 6. Conclusions

The elastic wave propagation in tetrachiral materials and metamaterials has been studied, according to the Floquet-Bloch theory for planar periodic systems. The periodic microstructure of the tetrachiral material is characterized by a square elementary cell, in which a central stiff ring is connected to four flexible ligaments. A parametric mechanical model has been formulated, by adopting a low-dimensional Lagrangian description of the microstructure as a square lattice of rigid annular bodies, connected by elastic unshearable beams.

The eigenproblem governing the spectral properties of the material has been solved through perturbation methods, suited to perform local sensitivity analyses of the dispersion functions. The local sensitivities to multi-parameter perturbations – in particular – have been assessed by applying a general algebraic formulation for the series expansion of a multivariable composite function up to a generic order. Therefore, as significant novelty with respect to other asymptotic approaches, a high-dimensional perturbation has been considered. Indeed, both the wavenumbers and the mechanical parameters are considered independent components of the same perturbation vector, spanning a small but multi-dimensional region of the parameter space. The closed form parametric solution of the perturbation equations has been achieved for the common case of a simple eigenvalues in the generating solution. Moreover, as substantial result, the dispersion functions have been asymptotically determined up to the fourth-order eigensensitivity. A satisfying approximation accuracy has been obtained over large ranges along the different directions spanned by the perturbation vector in the parameter space. Among the other directions, the highly-accurate performance of the asymptotic approximation along the triangular boundary of the first Brillouin zone (fixed the cellular microstructure) has been remarked.

Focusing on the tetrachiral material, the asymptotic approximations of the dispersion functions have been exploited to accomplish a twofold achievement. First, the existence conditions of a low-frequency stop band, ranging between the acoustic and the optical branches of the spectrum, have been analytically determined. From the physical viewpoint, the existence condition requires the rotational inertia of the rigid body to be sufficiently smaller than its mass, where the required smallness essentially depends on the beam slenderness. Second, a fourth-order parametric approximation of the stop bandwidth (if it exists) has been obtained. The parametric formulas clearly highlights how the stop bandwidth can be enlarged either by reducing the ligament slenderness or by decreasing the rotational inertia of the rings, whereas it is rather independent of the ring diameter.

Focusing on the tetrachiral metamaterial, it has been verified how the multi-parameter perturbation technique does not require any methodological adjustment to deal with the higher dimension of the perturbation vector. Indeed, if the inertial resonators are modeled as simple linear oscillators, their mechanical properties (mass and frequency) can be treated as new components of the perturbating vector, opening new directions in the enlarged parameter space. Furthermore, the dispersion functions has been approximated with satisfying accuracy, even in the low-frequency range of the metamaterial, which is featured by a high spectral density. In this respect, the paradigmatic case in which the resonator is tuned with the double acoustic frequency of the microstructure has been analyzed, due to its technical relevance in connection with the employment of metamaterials as mechanical filters. A fourth-order analytical approximation of the band gap amplitude opened by the inertial resonator has been achieved. The lowest order terms, consisting of simple explicit functions of the resonator parameters, have clearly highlighted the key-role of the resonator mass in enlarging the band gap amplitude. Moreover, the effects of a small detuning in the resonator frequency have been analyzed. In particular, overtuned resonators have been found to open larger band gaps than undertuned resonators.

Finally, the fourth order analytical approximations of the dispersion curves have been employed to state the inverse problem related the resonator design. The resonator mass and detuning required to open a target stop band in the metamaterial spectrum (with assigned bandwidth and center frequency), have been determined by solving a couple of nonlinear algebraic equations. Approximate analytical formulas for the design problem solution have been also obtained. The solutions have highlighted that a small overtuning of the resonator frequency is required to open a stop band around the desired center frequency.

**Acknowledgments**

The authors acknowledge financial support of the (MURST) Italian Department for University and Scientific and Technological Research in the framework of the research MIUR Prin15 project 2015LYYXA8, Multi-scale mechanical models for the design and optimization of micro-structured smart materials and metamaterials, coordinated by prof. A. Corigliano.

## AppendixA. Physical matrices

### AppendixA.1. Tetrachiral material

The non-null coefficients $M_{ij}$ and $K_{ij}$ ($i, j = 1, 2, 3$) of the 3-by-3 matrices $\mathbf{M}_a(\mathbf{p})$ and $\mathbf{K}_a(\mathbf{p}, \mathbf{b})$ governing the wave equation for the tetrachiral material read

$$M_{11} = M_{22} = 1, \qquad M_{33} = \chi^2 \qquad (A.1)$$
$$K_{11} = c_1 - c_2 \cos\beta_1 - c_3 \cos\beta_2$$
$$K_{22} = c_1 - c_3 \cos\beta_1 - c_2 \cos\beta_2$$
$$K_{33} = c_4 + c_5(\cos\beta_1 + \cos\beta_2)$$
$$K_{12} = 2c_6(\cos\beta_1 - \cos\beta_2)$$
$$K_{13} = -\mathrm{i}\,(c_6 \sin\beta_1 + c_7 \sin\beta_2)$$
$$K_{23} = \mathrm{i}\,(c_7 \sin\beta_1 - c_6 \sin\beta_2)$$

where the mass coefficients must be divided by $\omega_c^2$ if the normalization frequency $\omega_c$ is not unitary and

$$c_1 = 2\left((1-\delta^2)^2 + 12\varrho^2\right)(1-\delta^2)^{-5/2} \qquad (A.2)$$
$$c_2 = 2\left((1-\delta^2)^3 + 12\delta^2\varrho^2\right)(1-\delta^2)^{-5/2}$$
$$c_3 = 2\left(\delta^2(1-\delta^2) + 12\varrho^2\right)(1-\delta^2)^{-3/2}$$
$$c_4 = \left(\delta^2(1-\delta^2) + 16\varrho^2\right)(1-\delta^2)^{-3/2}$$
$$c_5 = \tfrac{1}{2}\left(\delta^2(1-\delta^2) + 8\varrho^2\right)(1-\delta^2)^{-3/2}$$
$$c_6 = \delta\left((1-\delta^2)^2 - 12\varrho^2\right)(1-\delta^2)^{-2}$$
$$c_7 = \left(\delta^2(1-\delta^2) + 12\varrho^2\right)(1-\delta^2)^{-3/2}$$

are fully geometrical auxiliary parameters. Note that $K_{ji}$ is the complex conjugate of $K_{ij}$ by virtue of the Hermitian property.

### AppendixA.2. Tetrachiral metamaterial

The 6-by-6 matrices $\mathbf{M}_a(\mathbf{p})$ and $\mathbf{K}_a(\mathbf{p}, \mathbf{b})$ governing the wave equation for the tetrachiral metamaterial read

$$\mathbf{M}_a = \begin{bmatrix} \mathbf{M}_s & \mathbf{O} \\ \mathbf{O} & \mathbf{M}_r \end{bmatrix},\quad \mathbf{K}_a = \begin{bmatrix} \mathbf{K}_s + \mathbf{K}_r & -\mathbf{K}_r \\ -\mathbf{K}_r & \mathbf{K}_r \end{bmatrix} \qquad (A.3)$$

where the 3-by-3 submatrices $\mathbf{M}_s$ and $\mathbf{K}_s$ account for the cell microstructure and coincide with the matrices governing the tetrachiral material (with the coefficients (A.1)). The 3-by-3 submatrices $\mathbf{M}_r$ and $\mathbf{K}_r$ refer to the local resonators and read

$$\mathbf{M}_r = \mathrm{diag}\left(\alpha^2, \alpha^2, \alpha^2\chi_r^2\right) \qquad (A.4)$$
$$\mathbf{K}_r = \mathrm{diag}\left(\alpha^2\gamma^2, \alpha^2\gamma^2, \alpha^2\gamma_\theta^2\chi_r^2\right)$$

where, again, the mass coefficients must be divided by $\omega_c^2$ if the normalization frequency $\omega_c$ is not unitary.

## AppendixB. Eigenvalue approximation

According to the MPPM, the characteristic equation governing the free wave propagation in the tetrachiral material has the approximate solution in $\epsilon$-power series

$$\lambda(\epsilon) = \lambda^\circ + \epsilon\lambda' + \epsilon^2\lambda'' + \ldots + \epsilon^n \lambda^{(n)} + \ldots \qquad (B.1)$$

The lowest eigenvalue sensitivities $\lambda^{(n)}$ up to $n = 2$ are reported in the following for selected reference points $\boldsymbol{\mu}^\circ = (\mathbf{p}^\circ, \mathbf{b}^\circ)$ of the parameter space, corresponding to the $\mathbf{b}^\circ$-vertices $B_1, B_2, B_3$ of the triangular $\mathcal{B}_1$-zone. It can be verified that the odd eigensensitivities ($n = 1, 3, \ldots$) at these $\boldsymbol{\mu}^\circ$-points are either null or $\mathbf{b}$-independent. In the asymptotic approximations this result is required to preserve the mathematical symmetry of the exact dispersion functions, which are even functions with respect to the $\mathbf{b}^\circ$-vertices $B_1, B_2, B_3$ of the $\mathcal{B}_1$-zone. All the eigensensitivities must be multiplied by $\omega_c^2$ if the normalization frequency $\omega_c$ is not unitary.

### AppendixB.1. Tetrachiral material in $B_1$

The vertex $B_1$ of the $\mathcal{B}_1$-zone is pointed by the wavevector $\mathbf{b}_1^\circ = (0, 0)$. The corresponding point $\boldsymbol{\mu}^\circ$ of the parameter space determines a double eigenvalue $\lambda_{1,2}^\circ$ ($m^\circ = 2$) and a single eigenvalue $\lambda_3^\circ$ ($m^\circ = 1$). After the reabsorption of the bookkeeping parameter $\epsilon$, the expansion coefficients read

- $\lambda_{1,2}^\circ = 0, \quad \lambda_3^\circ = 2\dfrac{d_1^2}{\chi^2}$ (B.2)

- $\lambda_1'' = \dfrac{d_2^2\Sigma_1^2 - R_1^4}{4d_1^2}, \quad \lambda_2'' = \dfrac{d_2^2\Sigma_1^2 + R_1^4}{4d_1^2}$ (B.3)

$$\lambda_3'' = \dfrac{\delta^2\Sigma_1^2}{2d_1^2} + \dfrac{4d_3^2\delta^2 - d_4^2\Sigma_1^2}{4\chi^2}$$

with first-order eigensensitivities $\lambda_{1,2,3}' = 0$, and the auxiliary geometrical parameters

$$d_1^2 = (\delta^2 + 12\varrho^2), \quad d_2^2 = (\delta^2 + 24\varrho^2), \qquad (B.4)$$
$$d_3^2 = (\delta^2 + 36\varrho^2), \quad d_4^2 = (\delta^2 + 8\varrho^2)$$

and the radical quantity

$$R_1^4 = \left[\left(\delta^2\Sigma_1^2 + 24\varrho^2\Delta_1^2\right)^2 - 96\varrho^2\delta^2\beta_2^2\Delta_1^2\right]^{1/2} \qquad (B.5)$$

have been introduced, with the squared wavenumber sum $\Sigma_1^2 = (\beta_1^2 + \beta_2^2)$ and difference $\Delta_1^2 = (\beta_1^2 - \beta_2^2)$.

### AppendixB.2. Tetrachiral material in $B_2$

The vertex $B_2$ of the $\mathcal{B}_1$-zone is pointed by the wavevector $\mathbf{b}_2^\circ = (\pi, 0)$. The corresponding point $\boldsymbol{\mu}^\circ$ of the parameter space determines three single eigenvalues $\lambda_1^\circ, \lambda_2^\circ, \lambda_3^\circ$ ($m^\circ = 1$). After the reabsorption of the bookkeeping parameter $\epsilon$, the expansion coefficients read

- $\lambda_1^\circ = 0, \quad \lambda_2^\circ = 4, \quad \lambda_3^\circ = \dfrac{d_5^2}{\chi^2}$ (B.6)

- $\lambda_1'' = 16\varrho^2\left(3 + \dfrac{\beta_2^2}{d_5^2}\right),$ (B.7)

$$\lambda_2'' = 2\delta^2 - (\beta_1 - \pi)^2\left(1 + \dfrac{\delta^2}{d_5^2 - 4\chi^2}\right),$$

$$\lambda_3'' = \dfrac{\delta^2\beta_2^2}{d_5^2} + \dfrac{\delta^2(\beta_1 - \pi)^2}{d_5^2 - 4\chi^2} + \dfrac{d_4^2\Delta_2^2 + 2\delta^2 d_6^2}{4\chi^2},$$

with first-order eigensensitivities $\lambda'_{1,2,3} = 0$, and the auxiliary geometrical parameters

$$d_5^2 = (\delta^2 + 16\varrho^2), \quad d_6^2 = (\delta^2 + 48\varrho^2) \tag{B.8}$$

have been introduced, with the squared wavenumber difference $\Delta_2^2 = ((\beta_1 - \pi)^2 - \beta_2^2)$.

*AppendixB.3. Tetrachiral material in $B_3$*

The vertex $B_1$ of the $\mathcal{B}_1$-zone corresponds to the wavevector $\mathbf{b}_3^\circ = (\pi, \pi)$. The corresponding point $\boldsymbol{\mu}^\circ$ of the parameter space determines a double eigenvalue $\lambda_{1,2}^\circ$ ($m^\circ = 2$) and a single eigenvalue $\lambda_3^\circ$ ($m^\circ = 1$). After the reabsorption of the bookkeeping parameter $\epsilon$, the expansion coefficients read

- $\lambda_{1,2}^\circ = 4, \quad \lambda_3^\circ = 8\dfrac{\varrho^2}{\chi^2}$ (B.9)

- $\lambda_1'' = 2d_2^2 - \dfrac{b_1^2 \Sigma_3^2 + R_2^4}{8b_2^2},$ (B.10)

  $\lambda_2'' = 2d_2^2 - \dfrac{b_1^2 \Sigma_3^2 - R_2^4}{8b_2^2},$

  $\lambda_3'' = \dfrac{\varrho^2}{2\chi^2 b_2^2}\left(b_1^2 \Sigma_3^2 + 24\delta^2 b_2^2\right)$

with first-order eigensensitivities $\lambda'_{1,2,3} = 0$, and the auxiliary mechanical parameters

$$b_1^2 = (\delta^2 + 8\varrho^2 - 4\chi^2), \quad b_2^2 = (2\varrho^2 - \chi^2), \tag{B.11}$$
$$b_3^2 = (\delta^2 + 4\varrho^2 - 2\chi^2),$$

and the radical quantity

$$R_2^4 = \left[\left(b_1^2 \Sigma_3^2\right)^2 - 32 b_2^2 b_3^2 \Pi_3^4\right]^{1/2} \tag{B.12}$$

have been introduced, with the squared wavenumber sum $\Sigma_3^2 = (\beta_1 - \pi)^2 + (\beta_2 - \pi)^2$ and product $\Pi_3^4 = (\beta_1 - \pi)^2 (\beta_2 - \pi)^2$.

*AppendixB.4. Tetrachiral metamaterial in $B_1$*

The vertex $B_1$ of the $\mathcal{B}_1$-zone is pointed by the wavevector $\mathbf{b}_1^\circ = (0, 0)$. If the resonator is tuned by setting $\gamma^\circ = 2$, the corresponding point $\boldsymbol{\mu}^\circ$ of the parameter space determines two double eigenvalue $\lambda_{1,2}^\circ$ and $\lambda_{3,4}^\circ$ ($m^\circ = 2$) and two single eigenvalues $\lambda_5^\circ$ and $\lambda_6^\circ$ ($m^\circ = 1$). Focus is made on the double eigenvalue $\lambda_{3,4}^\circ$, which defines the upper limit of the band gap $BG_1$. As long as the vertices of the $\mathcal{B}_1$-zone are selected as check points for the sensitivity analyses related to band gaps, a null $\mathbf{b}'$-perturbation can be assumed. After the reabsorption of the bookkeeping parameter $\epsilon$, the expansion coefficients read

- $\lambda_{3,4}^\circ = 4$ (B.13)
- $\lambda'_{3,4} = 4\sigma$ (B.14)
- $\lambda''_{3,4} = 4\alpha^2 + \sigma^2$ (B.15)
- $\lambda'''_{3,4} = 4\alpha^2 \sigma$ (B.16)
- $\lambda''''_{3,4} = \alpha^2 \sigma^2$ (B.17)

where it can be highlighted that the assumption $\mathbf{b}' = \mathbf{0}$ limits the small perturbation to the $\mathbf{b}$-space, where the eigenvalue preserves its double multiplicity at all the approximation orders. On the contrary, removing this assumption lets the double eigenvalues split into two close single eigenvalues at the fourth order.

*AppendixB.5. Tetrachiral metamaterial in $B_3$*

The vertex $B_3$ of the $\mathcal{B}_1$-zone is pointed by the wavevector $\mathbf{b}_1^\circ = (\pi, \pi)$. If the resonator is tuned by setting $\gamma^\circ = 2$, the corresponding point $\boldsymbol{\mu}^\circ$ of the parameter space determines a quadruple eigenvalue $\lambda_{1,2,3,4}^\circ$ and two single eigenvalues $\lambda_5^\circ$ and $\lambda_6^\circ$ ($m^\circ = 1$). Focus is made on the quadruple eigenvalue $\lambda_{1,2,3,4}^\circ$, which defines the lower limit of the band gap $BG_1$. As long as the vertices of the $\mathcal{B}_1$-zone are selected as check points for the sensitivity analyses related to band gaps, a null $\mathbf{b}'$-perturbation can be assumed. After the reabsorption of the bookkeeping parameter $\epsilon$, the expansion coefficients read

- $\lambda_{1,2,3,4}^\circ = 4$ (B.18)
- $\lambda'_{1,2} = 2(\sigma - S_1)$ (B.19)

  $\lambda'_{3,4} = 2(\sigma + S_1)$

- $\lambda''_{1,2} = d_2^2 + \tfrac{1}{2}S_1^2 + \dfrac{\sigma\left(2d_2^2 - (12\alpha^2 + \sigma^2)\right)}{2S_1}$ (B.20)

  $\lambda''_{3,4} = d_2^2 + \tfrac{1}{2}S_1^2 - \dfrac{\sigma\left(2d_2^2 - (12\alpha^2 + \sigma^2)\right)}{2S_1}$

- $\lambda'''_{1,4} = 2\alpha^2 \sigma - \dfrac{\alpha^2 \left(2d_2^2 + 4\alpha^2 + 3\sigma^2\right)^2}{4S_1^3}$ (B.21)

  $\lambda'''_{3,4} = 2\alpha^2 \sigma + \dfrac{\alpha^2 \left(2d_2^2 + 4\alpha^2 + 3\sigma^2\right)^2}{4S_1^3}$

- $\lambda''''_{1,2} = \sum_{i=0}^{6} \dfrac{\alpha^{2i}}{D_1}\Big(\sum_{j=0}^{5} c_{ij}\sigma^{2j} + S_1 \sum_{j=1}^{5} c_{ij}^* \sigma^{2j-1}\Big)$ (B.22)

  $\lambda''''_{3,4} = \sum_{i=0}^{6} \dfrac{\alpha^{2i}}{D_3}\Big(\sum_{j=0}^{5} c_{ij}\sigma^{2j} - S_1 \sum_{j=1}^{5} c_{ij}^* \sigma^{2j-1}\Big)$

where, if the second nondimensional frequency of the resonator is fixed ($\gamma_\theta = 4$) for the sake of simplicity, the denominators are

$$D_1 = 64 S_1^5 \Big(4b_4^2 S_1^3 + \sigma\left(4b_{36}^2 \alpha^2 + b_6^2 \sigma^2 - 18 b_2^2 d_2^2\right)\Big) \tag{B.23}$$

$$D_3 = 64 S_1^5 \Big(4b_4^2 S_1^3 - \sigma\left(4b_{36}^2 \alpha^2 + b_6^2 \sigma^2 - 18 b_2^2 d_2^2\right)\Big)$$

where the relevant quantity $S_1 = (4\alpha^2 + \sigma^2)^{1/2}$ depends only on the resonator parameters, while the coefficients $c_{ij}, c_{ij}^*$ and $b_4^2, b_6^2, b_{36}^2$ depend only on the mechanical parameters of the cell microstructure.

A complementary issue, inherent to perturbation techniques, is the discussion about the validity regions of the asymptotic approximation. Coarsely, the approximate eigensolution is not uniformly valid if certain small multi-parameter perturbations let the ratio between two consecutive terms of the approximation series become $O(\epsilon^0)$, or higher. Specifically, the asymptotic approximation can be expected to break down for the parameter combinations which let the higher eigensensitivity $\lambda^{(n)}$ increase up to quantitatively approach or overcome the lower eigensensitivity $\lambda^{(n-1)}$.

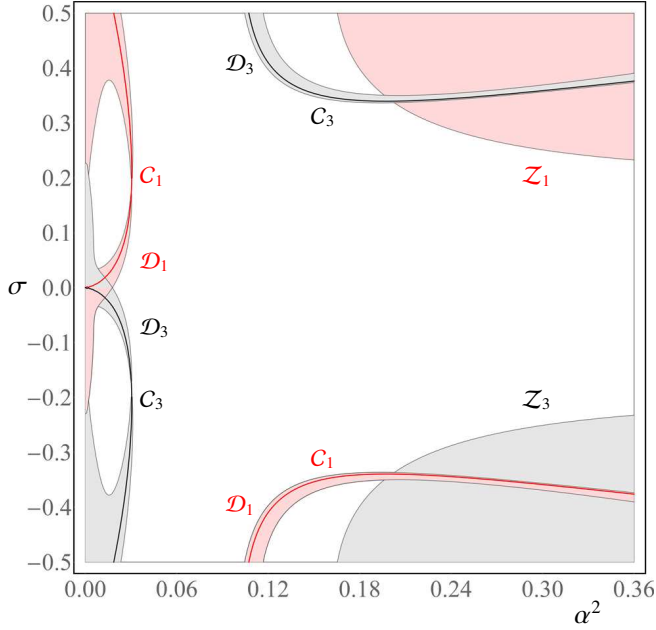

Figure B.9: Validity map of the asymptotic approximation of the eigenvalues in the $(\alpha^2, \sigma)$-space of the resonator parameters for fixed $(\delta, \varrho, \chi) = (1/10, 1/10, 1/12)$ and $(\gamma_\theta, \chi_r) = (4, 1/10)$.

Figure B.9 shows a validity map of the approximate eigensolution provided by the formulas (B.18)-(B.22) in the $(\alpha^2, \sigma)$-space of the resonator parameters. The two pink regions $\mathcal{D}_1$ collect all the $(\alpha^2, \sigma)$-points which do not satisfy the validity condition $\lambda''''_{1,2} \leq \frac{1}{3}\lambda'''_{1,2}$. Of course, the one-third ratio is largely conservative, but useful to graphically appreciate the $\mathcal{D}_1$-regions. Indeed, if the eigensensitivity ratio approaches unity, the $\mathcal{D}_1$-regions become narrower and narrower. In particular, the $\mathcal{D}_1$-boundaries become closer and closer to the red $\mathcal{C}_1$-curves, marking the $(\alpha^2, \sigma)$-combinations zeroing the denominator $D_1$ in the formula (B.23). The region $\mathcal{Z}_1$ preserves the approximation validity, since the $\mathcal{Z}_1$-points satisfies the inequality by virtue of very small or null $\lambda'''_{1,3}$-values. A similar discussion can be drawn for the gray $\mathcal{D}_3$-regions and the black $\mathcal{C}_3$-curves, which are related to the condition $\lambda''''_{3,4} \leq \frac{1}{3}\lambda'''_{3,4}$.

The non null coefficients of the sums in (B.23), depending only on the mechanical parameters of the cell microstructure, read

$$c_{60} = 3072 b_2^2 \tag{B.24}$$
$$c_{50} = 6144 b_2^2 d_2^2$$
$$c_{40} = -1536 \left(b_{10}^2 \delta^4 + \varrho^2 \left(b_{37}^2 \delta^2 - 1728 b_2^2 \varrho^2\right)\right)$$
$$c_{30} = -1536 b_2^2 d_2^2 \left(11 \delta^4 + 48 \varrho^2 d_8^2\right)$$

$$c_{20} = 192 b_2^2 d_2^8$$
$$c_{51} = -1024 b_8^2$$
$$c_{41} = -1536 b_5^2 d_2^2$$
$$c_{31} = 256 \left(b_{16}^2 \delta^4 + 48 \varrho^2 \left(b_{40}^2 \delta^2 + 12 \varrho^2 b_{47}^2\right)\right)$$
$$c_{21} = -128 d_2^2 \left(b_{24}^2 \delta^4 + 48 \varrho^2 \left(b_{46}^2 \delta^2 + 12 \varrho^2 b_{48}^2\right)\right)$$
$$c_{11} = 384 b_2^2 d_2^8$$
$$c_{42} = -128 b_{11}^2$$
$$c_{32} = 128 b_{14}^2 d_2^2$$
$$c_{22} = 32 \left(5 b_{20}^2 \delta^4 + 48 \varrho^2 \left(b_{43}^2 \delta^2 - 12 \varrho^2 b_{49}^2\right)\right)$$
$$c_{12} = -128 d_2^2 \left(b_{26}^2 \delta^4 + 48 \varrho^2 \left(b_{44}^2 \delta^2 - 12 \varrho^2 b_{27}^2\right)\right)$$
$$c_{33} = -64 b_7^2$$
$$c_{23} = -32 b_{23}^2 d_2^2$$
$$c_{13} = 64 \left(b_{32}^2 \delta^4 + 48 \varrho^2 \left(b_{45}^2 \delta^2 + 12 \varrho^2 b_{50}^2\right)\right)$$
$$c_{03} = -864 b_2^2 d_2^2 d_7^2 \delta^2$$
$$c_{24} = -4 b_{19}^2$$
$$c_{14} = -32 b_{25}^2 d_2^2$$
$$c_{04} = 432 b_2^2 d_7^2 \delta^2$$
$$c_{15} = -72 b_{31}^2$$
$$c_{51}^* = -20480 b_4^2$$
$$c_{41}^* = -6144 b_4^2 d_2^2$$
$$c_{31}^* = 1024 \left(b_{15}^2 \delta^4 + 48 \varrho^2 \left(b_{38}^2 \delta^2 + 12 b_4^2 \varrho^2\right)\right)$$
$$c_{21}^* = -512 d_2^2 \left(b_{22}^2 \delta^4 + 48 \varrho^2 \left(b_{39}^2 \delta^2 + 12 b_4^2 \varrho^2\right)\right)$$
$$c_{42}^* = 2048 b_9^2$$
$$c_{32}^* = -512 b_{13}^2 d_2^2$$
$$c_{22}^* = 256 \left(b_{17}^2 \delta^4 + 48 \varrho^2 \left(b_{46}^2 \delta^2 - 48 b_4^2 \varrho^2\right)\right)$$
$$c_{12}^* = -128 d_2^2 \left(b_{35}^2 \delta^4 + 48 \varrho^2 \left(b_{41}^2 \delta^2 + 12 b_4^2 \varrho^2\right)\right)$$
$$c_{33}^* = 512 b_{12}^2$$
$$c_{23}^* = -128 b_{21}^2 d_2^2$$
$$c_{13}^* = 320 \left(b_{30}^2 \delta^4 + 48 \varrho^2 \left(b_{42}^2 \delta^2 - 12 b_4 \varrho^2\right)\right)$$
$$c_{03}^* = -864 b_2^2 d_2^2 d_7^2 \delta^2$$
$$c_{24}^* = 128 b_{18}^2$$
$$c_{14}^* = -32 b_{28}^2 d_2^2$$
$$c_{04}^* = 432 b_2^2 d_7^2 \delta^2$$
$$c_{15}^* = 48 b_{29}^2$$

where the following auxiliary mechanical parameters have been introduced

$$b_4^2 = (\varrho^2 + \chi^2), \quad b_5^2 = (10 \varrho^2 - 11 \chi^2), \tag{B.25}$$
$$b_6^2 = (14 \varrho^2 - 13 \chi^2), \quad b_7^2 = (550 \varrho^2 - 683 \chi^2),$$
$$b_8^2 = (254 \varrho^2 - 205 \chi^2), \quad b_9^2 = (41 \varrho^2 - 40 \chi^2),$$
$$b_{10}^2 = (10 \varrho^2 - 53 \chi^2), \quad b_{11}^2 = (1454 \varrho^2 - 1327 \chi^2),$$

$b_{12}^2 = (104\varrho^2 - 85\chi^2),$ $\quad b_{13}^2 = (41\varrho^2 - 13\chi^2),$
$b_{14}^2 = (86\varrho^2 - 13\chi^2)$ $\quad b_{15}^2 = (163\varrho^2 - 80\chi^2),$
$b_{16}^2 = (554\varrho^2 - 67\chi^2),$ $\quad b_{17}^2 = (374\varrho^2 - 193\chi^2),$
$b_{18}^2 = (71\varrho^2 - 64\chi^2),$ $\quad b_{19}^2 = (1274\varrho^2 - 1741\chi^2),$
$b_{20}^2 = (442\varrho^2 - 125\chi^2),$ $\quad b_{21}^2 = (85\varrho^2 - 23\chi^2),$
$b_{22}^2 = (55\varrho^2 - 26\chi^2),$ $\quad b_{23}^2 = (266\varrho^2 - 211\chi^2),$
$b_{24}^2 = (386\varrho^2 - 199\chi^2),$ $\quad b_{25}^2 = (13\varrho^2 - 23\chi^2),$
$b_{26}^2 = (113\varrho^2 - 58\chi^2),$ $\quad b_{27}^2 = (13\varrho^2 - 5\chi^2),$
$b_{28}^2 = (47\varrho^2 - 7\chi^2),$ $\quad b_{29}^2 = (7\varrho^2 - 11\chi^2),$
$b_{30}^2 = (53\varrho^2 - 28\chi^2),$ $\quad b_{31}^2 = (4\varrho^2 - 7\chi^2),$
$b_{32}^2 = (251\varrho^2 - 100\chi^2),$ $\quad b_{33}^2 = (626\varrho^2 - 319\chi^2),$
$b_{34}^2 = (94\varrho^2 - 95\chi^2),$ $\quad b_{35}^2 = (109\varrho^2 - 53\chi^2),$
$b_{36}^2 = (50\varrho^2 - 31\chi^2),$ $\quad b_{37}^2 = (992\varrho^2 - 4336\chi^2),$
$b_{38}^2 = (271\varrho^2 - 134\chi^2),$ $\quad b_{39}^2 = (91\varrho^2 - 44\chi^2),$
$b_{40}^2 = (842\varrho^2 - 67\chi^2),$ $\quad b_{41}^2 = (181\varrho^2 - 89\chi^2),$
$b_{42}^2 = (89\varrho^2 - 46\chi^2),$ $\quad b_{43}^2 = (3746\varrho^2 - 1105\chi^2),$
$b_{44}^2 = (197\varrho^2 - 100\chi^2),$ $\quad b_{45}^2 = (403\varrho^2 - 164\chi^2),$
$b_{46}^2 = (626\varrho^2 - 319\chi^2),$ $\quad b_{47}^2 = (122\varrho^2 - 67\chi^2),$
$b_{48}^2 = (26\varrho^2 - 19\chi^2),$ $\quad b_{49}^2 = (94\varrho^2 - 95\chi^2),$
$b_{50}^2 = (23\varrho^2 - 4\chi^2),$
$d_7^2 = (\delta^2 + 80\varrho^2),$ $\quad d_8^2 = (19\delta^2 - 12\varrho^2)$